\documentclass[a4paper,10 pt, openodd, a4paper]{article}
\usepackage{epsfig,amsmath,amssymb,graphics,color,calc,cite,psfrag}

\usepackage{amsmath} 
\usepackage{amsthm}
\usepackage{amssymb}
\usepackage{amsopn}
\newtheorem{teo}{Theorem}[section]      \newtheorem{pro}[teo]{Proposition}
\newtheorem{defi}[teo]{Definition}      \newtheorem{lem}[teo]{Lemma}
\newtheorem{cor}[teo]{Corollary}        \newtheorem{rem}[teo]{Remark}
\newtheorem{con}[teo]{Condition}
\newcommand{\bteo}[1]{\begin{teo}\label{#1}}
\newcommand{\bpro}[1]{\begin{pro}\label{#1}}
\newcommand{\bdefi}[1]{\begin{defi}\label{#1}}
\newcommand{\blem}[1]{\begin{lem}\label{#1}}
\newcommand{\bcor}[1]{\begin{cor}\label{#1}}
\newcommand{\brem}[1]{\begin{rem}\label{#1}}
\newcommand{\bcon}[1]{\begin{con}\label{#1}}

\newcommand{\eteo}{\end{teo}}   \newcommand{\epro}{\end{pro}}
\newcommand{\edefi}{\end{defi}} \newcommand{\elem}{\end{lem}}
\newcommand{\ecor}{\end{cor}}   \newcommand{\erem}{\end{rem}}
\newcommand{\econ}{\end{con}}

\renewcommand{\eqref}[1]{(\ref{#1})}

\newcommand{\besn}{\begin{equation*}}
\newcommand{\beasn}{\begin{eqnarray*}}

\renewcommand{\(}{\left(}               \renewcommand{\)}{\right)}

      \newcommand{\id}{{1 \mskip -5mu {\rm I}}}
 
         \newcommand{\h}{\eta}

       \renewcommand{\r}{\rho}

         \renewcommand{\L}{\Lambda}
        \renewcommand{\O}{\Omega}

    \newcommand{\cL}{\mathcal L}

     \newcommand{\bR}{\mathbb R}

     \newcommand{\bZ}{\mathbb Z} 
\newcommand{\be}{\begin{equation}}
\newcommand{\ee}{\end{equation}}
\newcommand{\bea}{\begin{eqnarray}}
\newcommand{\eea}{\end{eqnarray}}
\newcommand{\ba}[1]{\begin{array}{*{#1}{c}}}
\newcommand{\ea}{\end{array}}

\begin{document}

\title{Cooperative Behavior of Kinetically Constrained Lattice Gas Models of Glassy Dynamics}

\author{Cristina Toninelli\thanks{Laboratoire de Physique
    Th{\'e}orique de l'ENS, 24 rue Lhomond 75231 Paris Cedex, FRANCE},
Giulio Biroli\thanks{Service de Physique Th{\'e}orique, CEA/Saclay-Orme des Merisiers,
F-91191 Gif-sur-Yvette Cedex, FRANCE} and Daniel
S. Fisher\thanks{Lyman Laboratory of Physics, Harvard University, Cambridge, MA 02138, USA} }

\maketitle
\begin{abstract}
Kinetically constrained lattice models of
glasses introduced by Kob and Andersen (KA) are analyzed. It is proved that only two
behaviors are possible on hypercubic lattices:  either ergodicity at all densities or trivial non-ergodicity, depending on
the constraint parameter  and the dimensionality. But in the ergodic cases, the dynamics is shown to be
intrinsically cooperative at high densities giving rise to  glassy dynamics as observed in simulations. The cooperativity is characterized by two length scales whose behavior controls finite-size effects: these are essential for interpreting simulations.
In contrast to hypercubic lattices,  on Bethe lattices KA models undergo a dynamical
(jamming) phase transition at a critical density: this is characterized by diverging time and length scales and a discontinuous jump in the long-time limit of the density autocorrelation function. By analyzing  generalized Bethe lattices (with loops) that interpolate between 
hypercubic lattices and standard Bethe lattices, the crossover between the dynamical
transition that exists on these lattices and its absence in the hypercubic lattice limit is explored. Contact with earlier results are made via  analysis of the related  Fredrickson-Andersen models, followed by brief discussions of universality, of  other approaches to glass
transitions, and of some issues relevant for 
experiments.
\end{abstract}

\section{\bf Introduction}\label{introduction}

The glass transition --- whether a true transition or a sharp crossover in the dynamics --- is one of the  longest standing unsolved puzzles  in condensed matter physics. 
Many liquids, when cooled fast enough to avoid crystallization, initially remain liquid but at lower temperatures appear to freeze into solid-like
structures that nevertheless show no signs of crystalline order. The  time scales for
structural relaxation in such metastable {\it super-cooled} regimes  increase dramatically as the temperature is
lowered  and materials are  conventionally called glasses when the
structural relaxation times are longer than those accessible in  typical experiments.  
In {\it fragile liquids} the relaxation time $\tau$ increases
more rapidly than Arrhenius \cite{DeBenedettiStillinger,DeBenedetti} 
and  can be characterized by an   effective activation free energy 
 $d\log \tau/ d(1/T)$ that increases by as much as a factor of twenty over a
 narrow temperature range. \cite{DeBenedetti}
Moreover, near  glass ``transitions", relaxation processes typically become
complicated \cite{DeBenedettiStillinger}, often characterized by stretched
exponential decay  of temporal correlations. Concomitantly, persistent spatial
heterogeneities appear  in both experiments \cite{Erdiger,Weitz} and numerical
simulations \cite{Harrowell,Onuki,Glotzer,Berthier1}.

Yet, in spite of a great deal of theoretical effort over the past few
decades, a real understanding of these and related phenomena is still
lacking. Indeed,  the most basic issues are still
unresolved: Is the rapid slowing down due to proximity to an
equilibrium phase transition of some kind (albeit perhaps in a
restricted part of phase space)? Or is the underlying cause entirely
dynamical? The latter scenario is supported by the lack of evidence for a
diverging static correlation length and the weak
 temperature dependence of structural correlations,
\cite{DeBenedetti} while growing 
dynamical correlation lengths have been found  in both experiments and simulations. \cite{Weitz,Erdiger,Harrowell,Onuki,Glotzer,Berthier1}
Whether or not there is an actual transition of some kind, what is the
physical mechanism responsible for the slowing down of the dynamics
and the growth of dynamical length scales?\\
%And if so, how might one probe these, even in principle?\\
%I would skip this questions because there are already some claims in
%the literature on how measure it

Theoretical progress has been hampered by a shortage of models that
capture at least some of the features believed to be essential, yet
are simple enough to analyze. In other fields of statistical physics
simple models have played vital roles leading  to new understanding, to sharpening of questions, and to development of improved
models.

\subsection{Kinetically constrained lattice models} 
\label{kin-constr-models}

One of the few types of simple models for glasses are the {\it kinetically constrained} models, in particular, lattice models with constrained dynamics; for reviews see \cite{Jackle,ReviewKLG}.   In this paper, we focus on a particularly simple class of models introduced by
 Kob and Andersen (KA) \cite{KA}. Before introducing these KA models and presenting our results,
 we give a short overview of  kinetically constrained models more generally, including their motivation  and the key questions about their behavior that could  provide a deeper understanding of glass transitions in real systems.

The basic Ansatz that motivates kinetically constrained models (see
 e.g. \cite{ChandlerGarrahanKLG}) is that the key to glass transitions is the
 geometrical constraints on  rearrangements of the atoms or  molecules and the
 effects of these constraints on the dynamics; static correlations, beyond
 those present in dense liquids, are assumed to play no role. The simplest
 models, which include KA models, are stochastic lattice gases with no static interactions beyond a hard core which restricts each site to be occupied by at most a single particle. The non-trivial aspect of these models arises from {\it local constraints} that are imposed on the motion of particles. Typically, the dynamics is given by
 a continuous time Markov process with
 each particle attempting, at a fixed
 rate, to jump to a randomly chosen empty neighboring site, but such a move is
 allowed only if the local configuration satisfies one or more constraints
 that depend on occupations of other nearby sites. 

The motivation for the kinetic constraints in conservative models
arises from the behavior of molecules in dense liquids:  
the presence of surrounding particles can severely inhibit the motion of a
molecule. In particular, a
molecule can be  {\it caged} by its neighbors in such a way as to stop 
it moving a substantial distance until the cage is opened by the  motion of
other particles. When caging is ubiquitous, such local constraints can produce a degree of cooperative behavior that slows
down the dynamics: one can readily imagine
this to be the underlying mechanism that causes glass transitions.
Since the idea behind the caging picture is that glass transitions are purely dynamical phenomena,
with any changes in static correlations playing a minor role (see
e.g. \cite{ChandlerGarrahanKLG}), 
it is natural to explore models whose dynamics satisfy detailed
balance with respect to some simple ensemble, in particular the trivial measure that  is uniform  over
all configurations with the same number of particles: this corresponds to ignoring static interactions beyond 
hard cores. [These models are therefore
quite different from the {\it statically} constrained "lattice glass models" introduced in
\cite{BiroliMezard}]. Because of the neglect of interactions beyond hard cores, no equilibrium transition can take place in purely kinetically constrained models; clearly they cannot approximate the behavior
of glass-forming liquids in all regimes, including the supercooling that avoids crystallization. Nevertheless, despite
their simplicity and discrete character, kinetically constrained lattice models might still capture, at least on some range of time scales,  the key dynamical aspects of real glass transitions.  Indeed,  numerical simulations show that some such models do display sluggish dynamics
and much phenomenology that is reminiscent of glassy
behavior and its onset as density increases \cite{ReviewKLG,Jackle}.  There
has been a renewal of interest in these models 
for explaining the behavior of glass-forming liquids
\cite{RecentKLG1,RecentKLG2,RecentKLG3,RecentKLG4,Berthier,Pitard,ChandlerGarrahanKLG,Whitelam,KurchanJ,Kurchan,Barrat,FranzKA,decoupling}. 
Kinetically constrained lattice gases have also
been studied in the context of granular systems \cite{Sellitto} which also display a glass-like dynamical arrest
as their density is increased; this so-called
{\it jamming transition} \cite{Jaeger, Liu} occurs at a density well below close--packing
density and can thus not have purely entropic origin. 

Note that there is another class of kinetically constrained models whose elementary degrees
of freedom are spins and not particles. An important example on which we will focus on section \ref{FA} 
is provided by the  Fredrickson and Andersen (FA) models \cite{FA}.   
These are the non-conservative version of the previous ones: there is no 
static interaction among spins on different sites and the non-trivial aspect arises 
from the local constraints imposed on the possible spin flips. In this case, 
the motivation for kinetic constraints comes from the concept of {\sl dynamic facilitation} \cite{FA,ChandlerGarrahanKLG}:
the evidence of dynamical heterogeneities, i.e. the existence regions of space with very different relaxation times 
suggests that, on a coarse grained level, supercooled liquids might be considered
 as a mixture of mobile and non-mobile regions. The non-conservative character
comes from the fact that the ratio of mobile to non-mobile regions is  non-conserved in time. 
Kinetic constraints arise from the fact that the vicinity of mobile
regions is required in order to enable ({\sl facilitate}) mobility in a
non-mobile regions.

\subsection{Characterization of possible glass transitions}

The most crucial questions about kinetically constrained models of
glasses  are whether  dynamical arrest takes place with increasing density, and, if so, whether this is the result
of some type of sharp transition or instead a gradual freezing. In either case,
one needs to understand the cooperative nature of the dynamics: in
particular, the emergence of growing length scales that characterize the cooperativity. 

We briefly sketch several possible scenarios for glass transitions.
The most striking possibility would
be an actual {\it ergodic / non-ergodic transition} at some critical density:
this possibility, which corresponds to an analogous conjecture for the nature of the dynamical arrest in glass forming liquids, is the primary  subject of the present paper. Because of the simple nature of the KA models, this question can be completely separated from possible changes in  equilibrium static behavior.
As mentioned above, for kinetically constrained models whose dynamics conserves the number of particles and satisfies detailed balance with respect to the uniform (non-interacting)
measure, the trivial distribution that is flat over all configurations with the same density 
is always stationary and there can be no equilibrium transition.
Nevertheless, due to the constraints on the allowed particle moves, ergodicity could be broken.  Non-ergodicity means that even in the limit of long times, time averages of local quantities would {\it not} converge to averages over the static equilibrium ensemble.
This certainly occurs for  some kinetically constrained models on {\it finite} lattices for which  the configuration space typically breaks into
{\it disconnected irreducible components}: i.e. there exist two or more sets of configurations that can not be connected  to one another by any sequence of allowed moves.
A key question  is whether this non-ergodicity can persist in the thermodynamic limit for which the grand canonical ensemble at fugacity $\rho$ (Bernoulli product measure at density $\rho$) is an equilibrium distribution. 
% dobbiamo dire qualcosa di piu' su cosa e' ergodicita' qui?
Specifically, can an ergodic/non--ergodic transition
occur at some density $\rho_c$ such that the dynamics on the
infinite lattice is ergodic for $\rho<\rho_c$, but not ergodic for
$\rho>\rho_c$?  Such an ergodicity-breaking transition is often considered an {\it ideal glass
transition}; it has been advocated as the primary cause of the
slowing down of glass-forming liquids. Ergodicity breaking transitions have been found to occur in several approximate theories, 
including in mode coupling theory \cite{Gotze} and in fully-connected quenched random
spin models \cite{ReviewOE}.

A weaker type of  dynamical arrest could in principle take place even if ergodicity is not broken: in particular, a {\it diffusive/sub-diffusive transition} of the dynamics. Indeed, the motion of a {\it tracer particle} could become non-diffusive at high densities:
this could be observed by singling out of an infinite system that is in equilibrium, one particle, the tracer, and following its motion.  There exists a general rigorous result that restricts the behavior: as long as the dynamics is ergodic,  the position of  a  tagged particle converges under  diffusive space-time rescaling to  
Brownian motion with a density-dependent self-diffusion coefficient $D_S(\rho)\geq 0$ \cite{KV, S}.
But in order for the tagged particle to  diffuse,   the self-diffusion coefficient has to be non-zero:
for the normal hard core lattice gas (simple symmetric exclusion), this has
been proven \cite{KV} and in \cite{Sart} the result has been 
extended to all the models such that the rate of any given jump to an empty
neighbor is bounded away from zero. 
For kinetically constrained lattice gases, however, the tagged particle  {\it could} be  slowed enough that  $D_S$ becomes zero at sufficiently high density.
A diffusive/sub--diffusive transition would then presumably take place at some critical density, $\rho_D$ with $D_S>0$ for $\rho<\rho_D$ but $D_S=0$ for $\rho> \rho_D$. Such a transition has been conjectured to occur in real glasses: and several experiments on 
supercooled liquids near their
glass transition indeed show a striking 
decrease of the  mean square displacement of a tracer particle  over the observed range of times \cite{DeBenedetti}.\\

Another way in which diffusion could potentially break down involves the breakdown of
the conventional hydrodynamical limit  that usually holds on long length and time scales  \cite{S}. For lattice gases (and more generally for systems  with particle number conservation but no momentum conservation) the coarse-grained density profile evolves,  at long times, via the diffusion equation, $\partial_{t} \rho=\nabla \left( D(\rho)\nabla \rho\right)$, with the diffusivity, $D(\rho)$, dependent on the local density.
For generic kinetically constrained stochastic models, proving the existence of the hydrodynamic limit, as has been done for conventional hard core models, 
%including one with an overall particle flux,   
is not a trivial task. The natural conjecture, if a transition of some kind did occur, would be  the vanishing of the 
macroscopic diffusivity, $D(\rho)$, at high densities.  This would  give rise to a form of  {\it macroscopic arrest}, in which density profiles at high densities would evolve, if at all, only sub-diffusively.  Note that, in contrast to the diffusion of a tagged particle, evolution of  density profiles need not involve motion of individual particles far from their initial neighbors: density diffusion and self diffusion thus probe rather different aspects of the dynamics.

Consideration of the types of macroscopic diffusion raises the more general
question of relaxation towards equilibrium: this could be anomalously
slow even if the system remains ergodic at all densities. An
interesting quantity to consider is the dynamical structure factor, the
Fourier transform of the density-density correlation function at non-zero wave vector. For
normal lattice gases, the dynamical structure factor decays exponentially at long times but for kinetically constrained
models, the relaxation could perhaps be slower than exponential,
at least above some  critical density. Indeed in numerical simulations
of some kinetically constrained models at high density, sub-exponential
relaxation has been found over a range of more than three decades \cite{KA}. 
This, as well as the relaxation of different response functions (see e.g.
\cite{Harrowell2,Pitard}) is usually fitted with a
{\it stretched exponential} $S(t)\sim
\exp[-(t/\tau)^\beta]$, with $\beta<1$ (also called a
Kohlrausch-Williams-Watts function). 
Many  glass-forming liquids show similar behavior with 
relaxation functions  well-fitted by simple exponentials at high
temperatures, but  by stretched exponentials at low temperatures
\cite{Kol,KWW,DeBenedetti}.
In glass-forming liquids this behavior arises from  the superposition
of many relaxations with different decay rates, each corresponding
to  different spatial regions within a dynamically heterogeneous spatial structure which persists for longer time scales.
\cite{Erdiger}.
If such sub-exponential relaxation does occur in  kinetically constrained models --- even if only over a limited range of times --- a detailed understanding of it would shed light on the behavior of more
realistic
systems.
\footnote{It is important to note that stretched exponential relaxation can occur in the absence of such complications as heterogeneities:  indeed, the two dimensional Ising model with non-spin-conserving dynamics exhibits just such behavior in its ordered phase. \cite{HuseFisher-relaxn}}

Numerical simulations of 
Kob-Andersen models, as reviewed below, have  found
very sluggish dynamics at high density \cite{KA},  with much 
phenomenology that is very similar to glass-forming liquids.
Is this a signature of some type of real dynamical glass transition at a critical density? More generally, does one exists for at
least some versions of the KA models?  If not, what is the basic
mechanism for the extremely rapid onset of the slowing down as the  density is increased? Are there collective
processes involved? What if any, are the characteristic length scales,
and how are the time scales related to these? What is the physical
mechanism giving rise to the dynamical heterogeneities? And, most
importantly, what can one learn from the KA models about the behavior of
more realistic models?  In this paper we will address all of these, although the last, regrettably,  only to a limited extent. \\

 Some of the results of this paper have already been presented
in a short letter \cite{letter} and the proof of the positivity of the self-diffusion coefficient
for KA models was presented in \cite{SelfToninelliBiroli}; these both made
use of the results derived in the present paper.

\subsection{\bf KA models: definition and heuristics}
\label{definition}

Kob-Andersen models were introduced\cite{KA}, as mentioned above,
 to test the conjecture that {\it cage effects}  in liquids can induce a
 sudden onset of  dynamical arrest as the density increases thereby  being
 responsible for a glass transition in such systems. The conditions for particles to be able to move in KA models mimic those caused by the geometrical constraints imposed by surrounding particles on the motion of  molecules in dense liquids.
 
The Kob-Andersen models we consider are natural generalizations of
those of reference \cite{KA}: they are defined on  $d-$dimensional
hypercubic lattices $\Lambda\in\bZ^d $ 
with zero or one particles allowed per lattice site and  a parameter, $m$,   
that represents the set of allowed particle motions as explained below (for a more formal definition see section
\ref{irreducibility}).  Each particle {\it attempts}, at rate unity, to move 
to a randomly chosen one of its $z=2d$ nearest neighbor sites. But a move from a site $x$ to a neighboring site $y$ is allowed {\it only
if} site $y$ is empty {\it and} the particle has no more than $m$ occupied
neighbors both before and after the move. If any of these constraints is not
satisfied the particle remains at site $x$ (at least until its next attempt to
move). The possible values of $m$ range from $m=0,\dots,2d-1$, the upper limit
being one less than the coordination number, $z$ (due to the hard core constraint). In finite systems, one can consider a fixed number, $N$, of particles.  But in the infinite systems of primary interest, the particle number density, $\rho$, must be specified instead. 
 
 As the interesting behavior occurs at high densities --- $\rho$ close to one ---
 it is useful to reformulate
the rules in terms of  {\it vacancy } motion, i.e dynamics of the empty sites. As can readily be 
verified, the above constraints with parameter $m$ correspond to {\it vacancies} allowed to move
{\it only if} both the initial site before the move, {\it and} the  final site after the move, have {\it at least $s=z-m-1$
neighboring vacancies}. The hypercubic KA models are thus completely defined
 by the parameters $d$ and $m$, or equivalently, $d$ and $s$; we will usually
 use the latter. The special case $m=2d-1$ corresponds to $s=0$: this KA model
 thus corresponds exactly to conventional hard-core lattice gas dynamics
 (i.e. simple symmetric exclusion process). 

 For all values of $s$ the {\it equilibrium static properties} are
identical to those of the hard-core lattice gas model with no other
interactions, and are hence {\it trivial}. But the dynamics depends
radically on $s$.  For $s\ge d$, the dynamics is so constrained that
KA models are trivially non-ergodic at any $\rho\in (0,1]$. This can be
checked by noticing that any fully occupied hypercube of
 can never be broken up, therefore at any finite density there exists a finite
 fraction of particles which are forever blocked. We  thus focus on
$0<s<d$.  

Why might one expect  interesting dynamics for $s\geq 1$?
In contrast to the conventional lattice gas ($s=0$) for which
individual vacancies can move freely, for non-zero $s$ on hypercubic
lattices, there are {\it no finite mobile  clusters of vacancies} that
can move freely --- albeit together --- in an otherwise fully occupied
system. By this we mean that,  if the
surrounding lattice is completely filled, 
it is not possible to construct a a path of
 allowed nearest
neighbor moves through which one can shift the cluster of vacancies.
This follows from the existence of infinite frozen configurations which, as none of the particles in them can move at all,  impede the motion of any cluster of vacancies.  The simplest infinite frozen configuration is a fully occupied slab of particles  that is infinite in
$d-s$ of the lattice directions and has width two in the  other $s$
directions:  by construction, none of the particles in such a 
slab can move  with the KA dynamical rules with parameter $s$.
[In contrast to the absence of mobile clusters of vacancies on hypercubic lattices with non-zero $s$, on a two-dimensional triangular lattice with $s=1$,
isolated vacancies cannot move but neighboring pairs of vacancies can
move freely and equilibrate the system (see \cite{SelfToninelliBiroli} for a detailed and rigorous discussion).]  
 
  Although  for non-zero $s$ on hypercubic lattices there are no
finite mobile clusters of vacancies, neither are there finite frozen
clusters for $s<d$. This follows because a particle at a corner of any
fully occupied region that is surrounded by empty sites can move;  thus on an
infinite lattice there are {\it no finite frozen
clusters}. Nevertheless, there do exist {\it infinite frozen sets} of
particles such as   the completely filled 
slab defined above.
 Such simple frozen configurations have very little entropy ---- in fact, zero entropy density --- but a much larger set of frozen configurations exist that {\it does} have non-zero entropy density.  For example, for the simplest case, $s=1$ on a square lattice, in addition to configurations with a single two-wide frozen slab that is infinitely long, any configuration with a set of such slabs that each begin and end at  T-junctions with other such slabs will be frozen.  A crucial question about static configurations is thus whether, at sufficiently high densities, almost all infinite configurations contain an infinite --- and hence connected --- set of frozen particles. As we shall see, answering this is by no means trivial. For all the nontrivial cases, the 
above observations imply that the sluggish dynamics  at high densities
must be {\it intrinsically collective}, involving rearrangements of increasingly larger 
and larger regions as the density increases. 

\subsection{Previous numerical results}
 
Before introducing the analytic methods we will use, 
we briefly summarize the results of various numerical simulations of
the best studied KA model:  the three--dimensional case with
$s=2$. It has been suggested  that these simulations provide support
for the conjecture of an ergodic / non-ergodic transition at a non-trivial density, $\bar{\rho}$. 
In reference \cite{KA}  the
 self-diffusion coefficient of a tagged-particle was measured 
and the results found to  
  fit well a power law form  that vanished at a finite density: $D_s\propto
 (\rho-\bar\rho)^{\alpha}$ with
${\alpha}\simeq 3.1$ and 
 $\bar\rho\simeq 0.881$.  Approaching this same {\it apparent critical density} the rate of temporal relaxation of
  density-density correlations appeared
 to vanish as an inverse power of $\r-\bar\r$. Both results --- which
hold over roughly four decades of $D_{S}$ 
and two decades of $\r-\bar\r$ --- were
strongly suggestive of a dynamical glass transition at
$\rho=\bar\rho$. 

Furthermore, the asymptotic decay of the Fourier transform of density-density
correlation function was fitted with the form
$S(t)\simeq \exp [(-t/\tau)^{\beta}]$ 
with an exponent $\beta$ close to one (exponential decay)
 for low and intermediate densities and decreasing
 monotonically with $\rho$ (stretched exponential decay) at sufficiently high
 density (for $\rho>0.75$).

Later work found that other typical
features of ``glassy dynamics'' occur at high densities in this $d=3$,
$s=2$ model. 

 In \cite{FranzKA} numerical simulations found {\it dynamical heterogeneities} to be persistent for very long times, at densities higher than roughly  $\bar\rho$.
In \cite{Kurchan} the effects  of boundary sources of particle ---  corresponding to a grand canonical ensemble --- were studied, in particular by quenching the chemical potential $\mu$ of the sources below the apparent critical value that corresponds to the  equilibrium $\mu(\bar\rho)$.  A non-equilibrium regime was evident, and ``aging" effects similar to those observed in glasses were found.  Properties that play roles in various theoretical approaches to glass transitions have also been studied.
In particular, the relationship between the ``configurational entropy"
of the number of distinct frozen configurations and an ``effective temperature" which emerges from
the aging dynamics was investigated to test Edward's hypothesis 
\cite{Barrat} as formulated by Kurchan in the context of aging
dynamics;  this had been found in
certain mean field models studied in the context of glasses, see
\cite{KurchanJ}. Edwards' hypothesis  appears consistent with numerical
results on the $d=3,\  s=2$ KA model \cite{Barrat}.

The results announced in \cite{letter} and explained in detail in the following
provide a theoretical framework to interpret numerical simulations.
Two recent works \cite{Berthier,Pitard} have indeed analyzed
the $d=3,\  s=2$ KA model taking into account our results and checking
some of our analytic predictions.

\subsection{Summary of exact results}

From the numerical simulations described above, it is clear that the
dynamics of some KA models is very sluggish  at high density and
displays much of the basic  phenomenology of  glassy dynamics observed in many experimental systems.  Furthermore, there appears to be a rather sharp onset of the glassy behavior as the density is increased. 
 The analytical methods of this paper yield an understanding of the mechanisms underlying the slow dynamics of KA models and 
address the basic qualitative and quantitative questions about them.

The primary question  is whether there exists some kind of  {\it actual} dynamical
 transition for the $d=3$ $s=2$ (or other) KA model, or whether the {\it apparent} ``transition" seen in simulations is only a 
 result of the  finite size and finite time limitations of such simulations.  
  We first 
 investigate whether an ergodic/non--ergodic transition can occur at a non-trivial density in KA models.  We conclude that, for any dimension and any choice
 of the constraint parameter $s<d$, an {\it ergodic / non-ergodic transition cannot take
 place} at a non-trivial density $\rho\in(0,1)$. In other words, the
 sluggish behaviour found in  simulations cannot be the
 mark of an ``ideal" glass transition.  
More quantitatively, we derive a density
 dependent characteristic length scale $\Xi(\rho)$ which separates two
 different regimes:  for samples with linear dimensions $L\ll\Xi(\rho)$ the configuration space breaks
into many ergodic components; while for those with $L\gg\Xi(\rho)$ a single ergodic component 
 dominates --- although other disconnected components still exist (a precise
 definition of $\Xi(\rho)$ is given below). The length,
 $\Xi$, whose dependence  on $\r$ varies with $d$ and $s$, diverges at high density:
 $\Xi(\rho)\sim\exp\exp \left(c/(1-\rho)\right)$ for the $d=3$ and $s=2$ model.
 Therefore, large systems of size $L^3$, are likely to appear ergodic only  if
 $\rho<\rho_c(L)=1-\left(c/\log\log L\right)$. Although for smaller sizes
the apparent ergodicity breaking takes place at lower values of the density, the dependence on $L$ is very weak for large $L$. Of course, to properly analyze data from simulations,
 knowledge of the {\it existence} of a cross over length is a key ingredient for disentangling the 
finite size effects which will always be limiting at high densities.
 
The fact that ergodicity holds in infinite systems at all densities for $s<d$, does not rule out the possibility of a more subtle type of dynamic transition at a non-trivial density. In particular, a
diffusive/sub--diffusive transition, if it existed, would 
explain 
%the apparent divergence of typical relaxation times and 
the apparent vanishing of the self
diffusion coefficient found in the above-mentioned numerical
simulations \cite{KA}.  We have studied the asymptotic behaviour
of the diffusion coefficient, $D_S$, of a tagged particle. 
The result,  discussed in detail in \cite{letter} and
\cite{SelfToninelliBiroli}, is that $D_S$ is strictly positive at
any density $\rho<1$, so that  {\it a diffusive / sub--diffusive transition cannot take
place}. Moreover, this analysis \cite{letter} unveils the nature of the
collective processes which are needed for both equilibration and diffusion at high
density. The characteristic time scale, $\tau$, of the slow cooperative
dynamics can be calculated and for the non-trivial cases, namely $s<d$, $\tau$
diverges faster than any inverse power of $1-\rho$ as
$\rho\to 1$. 

Given these results for general KA models on hypercubic lattices, an immediate question arises as to whether KA models can {\it ever} exhibit a transition at a non-trivial density: if this can occur, it raises the possibility that the surprisingly sharp onset of the slow dynamics observed in simulations in, for example, $s=2,\ d=3$, might be due to the ``ghost" of such a transition. Experience suggests that the most likely models in which to find  actual transitions are ``mean-field-like" models \cite{ReviewOE}; thus we study KA models on tree structures, more precisely, Bethe lattices, which often provide realizations of mean field approximations.  We find that on Bethe lattices {\it there exists a critical density}, $\r_c$, above which ergodicity is broken and phase space breaks up into many disconnected components. The transition at $\r_c$  has aspects of both
first order and second order transitions.  
 
An interesting question that we leave open for future
investigation is the hydrodynamic behavior of  KA models on hypercubic lattices. Although 
we conjecture that on large enough length and time scales
hydrodynamics will be valid at any density less than one for ergodic KA models,
this might be very hard to  prove.  Assuming that the conjecture is correct, an important issue, related to both experiments
\cite{Erdigerpreprint} and simulations \cite{Berthier1} of
glass-forming liquids, is the characteristic length and time 
scale beyond which hydrodynamic behavior sets in, in particular, how these scales
increase with density.

A recent interesting analysis of this behavior within Kinetically Constrained
models have been presented in \cite{decoupling}.

%Of course another important  issue should be to establish the connection of KA model with real systems, either glasses or shaken granular media.

{\subsection{ Outline}

The organization of the remainder of this paper is as follows:
In section \ref{KAergodicity} we analyze $s<d$ KA models on hypercubic lattices
 and prove ergodicity  for any density $\rho<1$. First, in \ref{d2} we 
analyze the  simplest interesting case, $d=2$, $s=1$, and prove
that with unit probability there exists an
irreducible component of the configuration space 
in the thermodynamic limit and find an upper
bound, $\Xi^u(\rho)$, on the crossover length $\Xi(\rho)$. 
In  \ref{d3}, we extend these results to the case $d=3$, $s=2$ originally considered by
Kob and Andersen, and
in \ref{s1} derive analogous results  for $s=1$ in general $d$.
The convergence of large systems to the infinite system behavior is studied in  \ref{secexponential}; we prove that for $L>\Xi^u(\rho)$ the
probability of the maximal irreducible component goes to one at least
exponentially rapidly. Using these results, in  \ref{ds} we extend the irreducibility
proof and find upper bounds for the crossover
length for all  $s<d$.  
Finally, in \ref{irreducibility} we prove that irreducibility of the configuration space implies ergodicity.

Section \ref{Quantitative} is devoted to quantitative results on the
cross-over lengths. 
In \ref{bootstrap}, by using recent bootstrap percolation results to obtain lower bounds on $\Xi(\rho)$ and combining these with the upper bounds,
we find the asymptotic behavior of the density-dependence of the cross over length, up to an undetermined constant. In
section \ref{optimal}, we calculate
the exact value of this constant for the case $d=2$ $s=1$ 
via the identification of the dominant mechanism which restores
ergodicity in large systems at high density. 
In \ref{optimal3} we formulate an analogous conjecture for the dominant
high density behavior of the original KA model, $d=3$ $s=2$. 

In section \ref{picture} the physical
picture for the cooperative high density dynamics (which was presented 
in \cite{letter}) is explained and related to the results derived in this paper; details are left for a forthcoming paper.

In section \ref{sectionbethe}  KA models on Bethe lattices are studied.  After recalling
(section \ref{bethedefinition}) the definition and some properties of
Bethe lattices, in \ref{existence} and \ref{bethetransition} 
recursive relations are derived 
whose  fixed point solution implies 
that an ergodicity breaking
transition takes place at a non-trivial critical density $\rho_c$, for any $s<k$ --- the branching parameter of the Bethe lattice. The transition is shown to be discontinuous for
$s<k-1$ (section \ref{discontinuity}) but continuous for $s=k-1$ (section
\ref{continuity}). The existence and
location of this transition are related to a boostrap
percolation transition in \ref{betheboot}.  
In section
\ref{bethedynamics}, the focus is on quantitative results for Bethe lattice models:  we analyze in some detail the character of the transition
in the cases $k=3$, $s=1$ and $k=5$, $s=3$ and present results of
numerical simulations  for the  $k=3$,
$s=1$ model. The behaviour of the density--density
correlation function and the corresponding susceptibility are studied, along with the 
 configurational entropy, $S_c(\rho )$. By establishing a lower bound we prove
 that $S_c(\rho)$ jumps to a non-zero value
 at $\rho_c$.
We end this section, in
\ref{betheloops}, by  extending  to KA model on decorated Bethe
lattices: graphs with finite size loops that interpolate
between simple Bethe lattice and finite-dimensional 
hypercubic lattices.
  
In section \ref{FA} we analyze the related Fredrickson Andersen (FA)
models, proving, in \ref{FAergodicity}, ergodicity at any finite
temperature (completing an almost complete proof that had been given
by Fredrickson and Andersen themselves). In \ref{FAtime} we give
quantitative predictions on length and time scales for the $d=2$
two spin facilitated FA model.

Finally, in \ref{conclusions}, we draw conclusions and  discuss some possible connections  of our results  to other theoretical treatments of glass transitions. In particular, 
the analogy of KA models on Bethe lattices to  fully connected random p-spin
models.   We also comment briefly on possible extensions of KA models and connections to experiments.

\section{\bf Rigorous results for  KA models on hypercubic-lattices}
\label{KAergodicity}

In this section we analyze KA models on hypercubic lattices with the parameter $s$ in the non-trivial range  $1\le s\le d-1$. We prove that for any
 density $\rho<1$, on an infinite lattice such KA models are {\it ergodic}. 
In contrast, as explained earlier, for larger values of 
$s$, $s\ge d$, KA models on hypercubic lattices are non-ergodic for all positive $\r$ because of the existence of finite frozen clusters.
 There are thus only two possible behaviors for hypercubic lattices: either the KA process in the thermodynamic limit is ergodic
 at any density, i.e. $\rho_c=1$, or else it is never ergodic,
 i.e. $\rho_c=0$; therefore  {\it ergodic/non--ergodic transitions
 cannot take place in KA models on hypercubic lattices} for any $d$ and $s$.

The strategy of the proof is as follows. Let $\Lambda\in\bZ^d$ be an hypercube of
 linear size $L$. First we identify a component of the
configuration space $\O_\Lambda\equiv\{ 0,1\}^{|\L|}$ and show that it
is {\it irreducible}, i.e. any two different configurations belonging to this component can be connected  to each other by a sequence of elementary moves allowed by the KA rules. Second,  we prove that, with respect to the natural measure on the space of configurations, Bernoulli product
measure $\mu_{\L,\rho}$,  the
{\it probability of this single irreducible component goes to unity for $L\to \infty$}.  Finally we prove
that, thanks to the product form of Bernoulli measure, the {\it existence of an irreducible component with unit probability
implies ergodicity for the infinite system}, $\L=\bZ^d$. 

In the first four subsections, we give the first and second steps of the
proof for all choices of $d$ and $s$, leaving to 
\ref{irreducibility} the last step, which is proved by a general argument that is independent of  the specific parameters.

We start with the simplest non-trivial model, $d=2$, $s=1$,  giving the most details for this case.

\subsection{\bf Irreducibility for  d=2, s=1}
\label{d2}

The only non-trivial model on a square lattice is the case $s=1$: vacancies can move if and only if they have a neighboring vacancy both before and after the move.

\subsubsection{Frameable configurations}

In order to construct the large irreducible component of configuration space we start by defining a subset of the configurations in it. Specifically, 
 let us define {\it framed} configurations of any square or rectangle as those in which all the boundary sites are
 empty (see figure \ref{fra1}). We then define as  {\it frameable} any configuration from which, by an allowed sequence of
 elementary moves, a {\it framed} configuration can be reached.  As we show below, any two
 framed configurations that have the same number of particles can be connected to one another by a sequence
allowed moves; the same then applies to all frameable configurations by definition.  
%The irreducible sets of all frameable configurations of an $L^2$ square with fixed total (conserved)  particle number  form a basis of the irreducible component of the infinite system.

\subsubsection{Irreducibility of frameable set}

The irreducibility of a set of frameable configurations can be checked as
 follows. Note that even though it is not needed for the present purposes we shall prove irreducibility considering the particles as distinguishable.  This
result is important for the proof of the positivity of
the self-diffusion coefficient \cite{SelfToninelliBiroli}.
Consider a pair of neighboring sites $\{i,j\}$, with for example
particles A and B respectively on $\{i,j\}$ and $j$ immediately to the right of $i$.
\begin{figure}[bt]
\psfrag{A}[][]{$A$}
\psfrag{B}[][]{$B$}
\centerline{
\includegraphics[width=0.25\columnwidth]{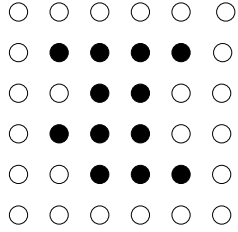}}
\caption{
A $6$ by $6$ framed configuration. Filled dots are occupied sites;
empty dots  vacant sites.}
%  In figure
%\ref{fra2}, \ref{fra3} and \ref{fra3bis} a sequence of moves is shown whose net result is
%the exchange of the neighboring particles labeled $A$ and $B$.} 
\label{fra1}
\end{figure}
 To prove the needed result, it is enough to show
 that, for any choice of the framed configuration, 
it is possible to perform the permutation of A with B
leaving at the end of the process all the other particles in their initial configuration.

 Starting from the bottom right corner  it is possible to raise the bottom row
 of holes  as shown in first line of figure \ref{fra2}, and likewise starting from the top
 right corner, to lower the top row of holes.  This procedure can be
 iterated until the row that contains sites $i$ and $j$ is
 ``sandwiched'' between two rows of holes (first configuration on second line
 of figure \ref{fra2}).
  At this point, using the surrounding vacancies, one can easily 
construct a path
 performing  
 the desired permutation of particle A and B from site $i$ to $j$ (second and
 third line of figure
 \ref{fra2}). Following this exchange,
 the initial configuration of the rest of the
 lattice can be restored by reversing the moves of the two rows of
holes after which all particles other than A and B have returned to their original sites. The full  procedure can be performed if instead there is a
particle and a hole on $\{i,j\}$.  Since any framed configuration can be
transformed to any other framed configuration with the same number of
 particles through a sequence of permutations of  nearest neighbor particles and/or
holes,  all the framed configurations with the same number of
 particles belong to the same irreducible component. By definition,  so also do
 all frameable configurations, a much larger set.
  
 %any two frameable configurations can be connected by a path which goes trhough the two corresponding framed configurations.

\begin{figure}[bt]
\psfrag{A}[][]{$A$}
\psfrag{B}[][]{$B$}
\centerline{ 
\includegraphics[width=\columnwidth]{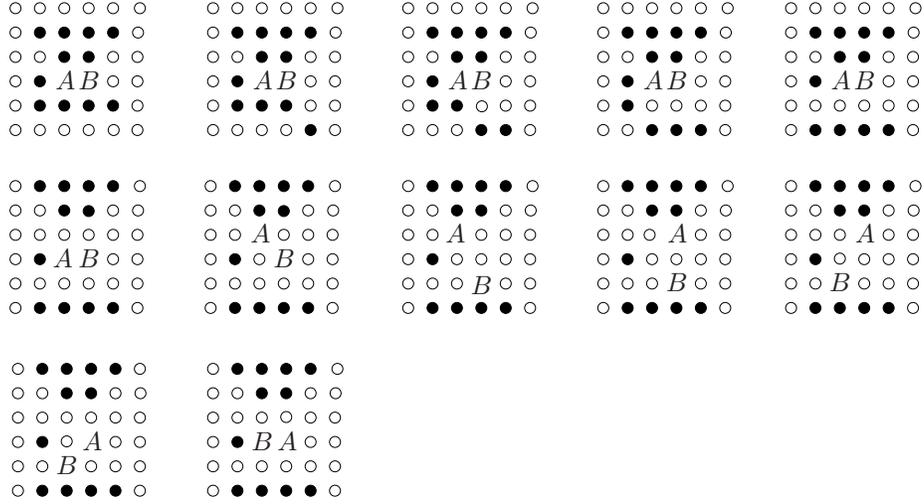}}
\caption{Sequence of allowed elementary moves starting from a generic 6 by 6
 framed  configuration, whose net result is the exchange of two neighbouring
  particles, A and B, inside the square. Figures in the first line
  show the elementary moves connecting the initial configuration
to a configuration with the bottom row of holes raised by
  one. By analogous moves (skipped in the figure) one can lower the top row of
  holes of two steps. Now (first figure of second row) particles $A$ and $B$  
are  ``sandwiched'' between two rows of holes. The remaining figures
 show how to exchange particles A and
  B starting from this configuration. Note that no other particles have been
  moved during this exchange. The configuration which is equal to the initial
  one (first line, first
  figure) on the sites not occupied by A and B 
can subsequently be restored by using sequences of moves
  analogous to those in the first line  to raise the one full row of holes and
  lower the other one.
\label{fra2}}
\end{figure}
%\begin{figure}[bt]
%\psfrag{A}[][]{$A$}
%\psfrag{B}[][]{$B$}
%\centerline{    
%\includegraphics[width=0.3\columnwidth]{fra3n.eps}}
%\caption{Configuration that can be reached from that of figure \ref{fra1}
%by the sequence shown in figure \ref{fra2} and analogous moves which lower the
%top row of holes twice.  
%\
%label{fra3}}
%\end{figure}

%\begin{figure}[bt]
%\psfrag{A}[][]{$A$}
%\psfrag{B}[][]{$B$}
%\centerline{ 
%\includegraphics[width=\columnwidth]{fra4n.eps}}
%\caption{Sequence of allowed elementary moves to exchange particles $A$ and $B$ starting form the configuration
%  in figure \ref{fra3}. 
%\label{fra3bis}}
%\end{figure}

\subsubsection{Frameable probability}

The next step is proving that, of all configurations with  density $\rho$, the fraction  that are frameable, $\mu_{\Lambda,\rho}({\cal{F}})$, approaches one in the limit $L\to \infty$. 
We will do this by first constructing a large subset of $\cal{F}$.

 Consider a four by four configuration which has  at its center a two by two
square of holes --- a two by two framed configuration ---  and in the surrounding shell at least two holes adjacent to each side of the inner square. It is easy
to check (see figure \ref{fra4}) that such a four by four configuration is
frameable. This procedure can be iterated to grow an $L$ by $L$ frameable configuration starting from a two by two nucleus of vacancies and requiring at least two vacancies in each side of each subsequent shell. Therefore, $\mu_{\Lambda,\rho}({\cal{F}})$ is bounded from below by
the probability, $\mu_{\Lambda,\rho}({\cal{F}}^0)$, of 
frameable configurations constructed with this procedure from a two by two nucleus of vacancies centered at the
origin:
\begin{equation}
\mu_{\Lambda,\rho}({\cal{F}})\geq \mu_{\Lambda,\rho}({\cal{F}}^0)=(1-\rho )^{4}\prod_{l=1}^{(L-2)/2}
(1-\rho ^{2l}-2l\rho ^{2l-1}(1-\rho))^{4} \ .
\end{equation}
Note the as the size of the shell grows, the probability that there are the requisite number of vacancies on each side increases until becoming close to unity for sufficiently large shells: specifically when $l \gg 1/(1-\r)$.
\begin{figure}[bt]
\centerline{    
\includegraphics[width=\columnwidth]{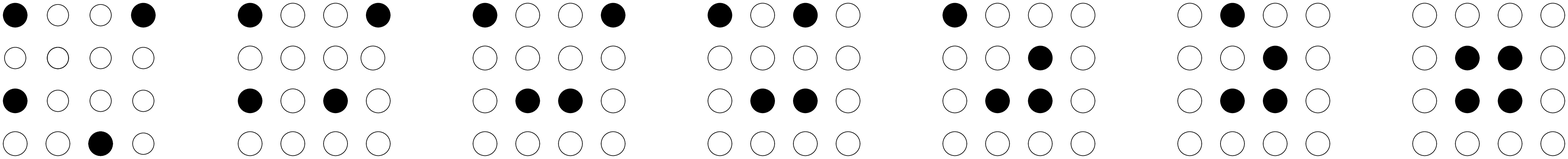}}
\caption{Growing $4$ by $4$ frameable configurations from a $2$ by $2$  framed configuration 
\label{fra4}}
\end{figure}
\begin{figure}[bt]
\centerline{   
\includegraphics[width=0.3\columnwidth]{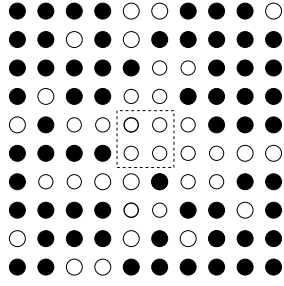}}
\caption{An $8$ by $8$ frameable configuration obtained from the growing procedure. The dashed square indicates the position of the  two by two seed. Note that the density in such configurations is small near the seed but  higher farther away.
\label{fra5}}
\end{figure}
The large $L$ behaviour of $\log \left(\mu_{\Lambda,\rho}({\cal{F}}^0)\right)$ is
determined by 
\be
\log (1-\rho)+\sum_{l}\log [1-\rho ^{2l}-2l\rho ^{2l-1}(1-\rho)]\simeq 
\log (1-\rho)-\sum_{l}[\rho ^{2l}+2l\rho ^{2l-1}(1-\rho)]\ ,
\label{prob1}
\ee
 which is a converging series for any $\rho <1$. Therefore $\mu_{\Lambda,\rho}({\cal{F}}^0)$ converges to a non-zero limit when $L\rightarrow \infty $ at fixed density.

Because of the divergence of the above series for $\rho=1$,  additional care is
required to analyze the behavior 
 when  both $\rho \rightarrow 1$ and $L\rightarrow \infty $. 
Let $a(\ell)\equiv\log (1-\rho ^{2\ell}-2\ell\rho ^{2\ell-1}(1-\rho))$. Since $a(\ell)$ is increasing in $\ell$, the following inequality holds:

\begin{eqnarray}
\label{inequ}
&&4a(1)+4\int_{1}^{\frac{L-2}{2}}d\ell\ a(\ell  \leq\nonumber \\
&&\log \mu_{\Lambda,\rho}({\cal{F}}^0) -4\log(1-\rho)  \leq\nonumber\\
&& 4\int_{2}^{\frac{L-2}{2}+1}d\ell \ a(\ell)
\end{eqnarray}
where the last term can be rewritten by using the change of variables 
$y=\rho^{2\ell}$ and expanding for $\rho\simeq 1$ as 
\begin{equation}
\label{integ}
\lim_{L\rightarrow \infty}\int_{2}^{\frac{L}{2}}\!\! d\ell \ a(\ell) \simeq\frac{1}{2(1-\rho) }\int_{0}^{1}
dy\frac{\log (1-y+y\log y)}{y}
\end{equation}
Therefore, by combining equations (\ref{inequ}) and (\ref{integ}) and using $\lim_{\rho\rightarrow 1}(1-\rho)\log (1-\rho)=0$
we get

\begin{equation}
\lim_{\rho\rightarrow 1}\lim_{L\rightarrow \infty} (1-\rho)\log \mu_{\Lambda,\rho}({\cal{F}}^0) =2\int_{0}^{1}
dy\frac{\log (1-y+y\log y)}{y}
\label{ris}
\end{equation} 
i.e.,  in the high density limit \footnote{Note that here and in the
following we will use for simplicity the loose notation 
$\mu_{\infty,\rho}({\cal{F}}^0)\simeq
 e^{-\(\frac{2c}{1-\rho }\)}$ although the correct mathematical 
expression is $\log \mu_{\infty,\rho}({\cal{F}}^0)\simeq
-\(\frac{2c}{1-\rho }\) $ up to vanishing corrections in the $\rho
\rightarrow 1$ limit.}

\begin{equation}
\label{muinf}
\lim_{L\to\infty}\mu_{\Lambda,\rho}({\cal{F}}^0)\equiv\mu_{\infty,\rho}({\cal{F}}^0)\simeq
 e^{-\(\frac{2c}{1-\rho }\)}
\end{equation}
with 

\begin{equation}
c=-\int_{0}^{1}
dy\frac{\log (1-y+y\log y)}{y}
\label{cc}
\end{equation}
Moreover from

\begin{eqnarray}
\log \mu_{\infty,\rho}({\cal{F}}^0)-\log \mu_{\Lambda,\rho}({\cal{F}}^0)
&\simeq&\frac{2}{1-\rho} \int_{1}^{\rho^{L}}dy\frac{\log (1-y+y\log y)}{y}\nonumber \\
&\simeq&\frac{2}{1-\rho}\int_{1}^{\rho^{L}}dy(-1+\log y)\nonumber\\
&=&\frac{2}{1-\rho}\left(\rho^{L}\log[\rho^{L}]-2\rho^{L}\right)\nonumber\\
&\simeq& \frac{-4\rho^{L}}{1-\rho}
\end{eqnarray}
we find that
$\mu_{\Lambda,\rho}({\cal{F}}^0)\simeq\mu_{\infty,\rho}({\cal{F}}^0)$
for
$L\gg\xi_{2,1}(\rho)$, defining a characteristic length, $\xi_{d,s}(\rho)$, with 
\begin{equation}
\label{xi2}
\xi_{2,1}= -\log (1-\rho)/(1-\rho) \ ,
\end{equation}
 which will play an important role in what follows.

We now need to get from the probability, $\mu_{\Lambda,\rho}({\cal{F}}^0)$, of a frameable region centered at the origin, to  the needed result:   the probability that  a square of size $L$ is frameable about {\it some} center, $\mu_{\Lambda,\rho}({\cal{F}})$,  converges to unity for $L\to \infty$.   To do this one must consider all the
 possible positions for the initial nucleus used in the growing procedure\footnote{The results obtained hold for the KA model on a finite square lattice with periodic boundary conditions. In this geometry, any point can be taken as the origin of the growing procedure.}.
Naively, we could simply make the argument that in a very large system, it is extremely unlikely that the system would not be  frameable around at least one of the many possible nuclei. But additional work is required to turn this argument into a proof since the events that the whole square is frameable starting from different nuclei are not independent.
We postpone the actual proof to the end of subsection \ref{s1}, where we consider the generic d-dimensional case, and here just outline the argument. 

The key idea is the following:
 from the above definition of $\xi$, the conditions needed to {\it further} expand a frameable square of size $L\gg\xi$ are satisfied with probability close to one.  Therefore, whether or not  one can grow frameable configurations starting from nuclei at the centers of {\it distinct} squares of size $\xi$ are
almost independent. By considering that there are $O\left(L^2/\xi^2\right)$
 disjoint squares of size $\xi$ inside the $L$ by $L$ square, $\L$, we
 conclude that $\mu_{\L,\rho}({\cal{F}})$ is almost one for $L\gg \xi
 /\sqrt{\mu_{\infty,\rho}({\cal{F}}^0)}\simeq \exp \left[c/(1-\rho)\right]$ and approaches one as $L\to \infty$.
The rigorous version of this argument given in \ref{s1} completes the proof of irreducibility in the thermodynamic limit.

Let us now turn to the analysis of finite size effects.
As stated above, the
probability of the irreducible frameable component goes to one when
$L\to\infty$ at any fixed density $\rho\in [0,1)$. On the other hand, for any
finite and fixed value of $L$, the probability of any irreducible component is
strictly smaller than one: blocked configurations exist and therefore the
configuration space is never covered by a single component. 
Furthermore, when $\rho\to 1$ the probability $\mu_{\L,\r}({\cal{M}})$
of the maximal irreducible
component (the irreducible component with the highest
probability) goes to zero since
blocked configurations are more and more important and each one is in a
different irreducible component. Therefore $\lim_{\rho\to
  1}\mu_{\L,\r}({\cal{M}})=0$ at fixed lattice size while  
 $\lim_{L\to\infty}\mu_{\L,\r}({\cal{M}})=1$ at fixed 
density. Thus the simultaneous $L\to \infty$ and
 $\rho\to 1 $ limit depends on the relationship between $L$ and $\rho$ as the limit is taken: this is typical of finite-size behavior near phase transitions. 
  %(and at least for $L$
 %sufficiently large the frameable component coincides with the maximal
% irreducible component). 
The
 crossover length $\Xi_{2,1}(\rho)$  is defined to separate the ``large" and ``small" system behaviors so that by
 sending $L\to\infty$ and $\rho\to 1$ with $L\gg\Xi_{2,1}(\rho)$ or
 $L\ll\Xi_{2,1}(\rho)$, respectively, one obtains $\mu_{\L,\r}({\cal{M}})\to 1$ 
or $\mu_{\L,\r}({\cal{M}})\to 0$, respectively. This defines the limiting behavior of $\Xi(\rho)$ up to an overall constant factor. A more precise definition is possible, for example, by  the sequence of $(L,\rho)$ pairs for which $\mu_{\L,\r}({\cal{M}})=\frac{1}{2}$.
Above results on the probability of the frameable component establish
therefore 
an upper bound on the crossover length:

\begin{equation}
\label{u2}
\Xi_{2,1}(\rho)\leq\Xi^u_{2,1}(\rho)\approx \exp \left[ \frac{c}{(1-\rho) }\right] \ ,
\end{equation} 
with $c$ defined in (\ref{cc}).
The reason why $\Xi^u$ is only an upper bound for the cross over length is
because while our framing argument guarantees that for $L>\Xi^u$  a single
irreducible component (the frameable one) dominates, it says nothing about
the probability of the {\it maximal} irreducible component, which could
already be close to unity for much smaller sizes. Indeed, if we have not got
roughly the correct size-dependent conditions for frameability, $L\times L$
squares could  be almost ergodic even for $(1-\rho)\log L\ll (1-\rho)\log ~\Xi^u$.  This
possibility is ruled out in section \ref{bootstrap} by establishing a lower
bound $\Xi^l$ that has  similar density dependence to $\Xi^u$, ensuring that
we do indeed have the correct asymptotic form, albeit with the wrong
coefficient $c$.
The exact coefficient will be calculated in
section \ref{optimal} 
thanks to the definition of a different framing procedure.

The cross-over length, $\Xi$, is also the scale above which frozen sets of particles are unlikely to exist.  In particular, a spanning  network of fully-occupied two-wide slabs that all start and end either at system boundaries or at T-junctions with other slabs --- a frozen configuration --- is likely to exist for squares of size $L<\Xi$, but unlikely to exist for much larger squares. How does this occur? In particular, if a number of squares of size somewhat smaller than $\Xi$ are put together to make one of size a few times $\Xi$, what happens to the (formerly) frozen bars that terminated inside one of the original squares?  A crucial feature is the {\it fragility} of a typical frozen network of slabs: if one slab is cut, as will occur if it ends on an edge of one of the now-interior smaller squares, this can trigger a catastrophic failure of most of the network. Such extreme fragility is why the set of all frozen configurations has measure zero in infinite systems in spite of its non-zero entropy density.

Before moving on to more complicated KA models, it is important to note that the whole proof of irreducibility is based on two ingredients: all configurations that contain a special frame of holes belong to the same irreducible component; and the frame can be created by starting from a small nucleus of vacancies and expanding to larger sizes by satisfying at each step  requirements that becomes less and less severe probabilistically. Therefore, the basic result for the simple square lattice model we have considered thus far is not strongly dependent on the exact set of dynamical constraints: in the  next subsections we show how it can be adapted to all  KA models on hypercubic lattices.

\subsubsection{Exponential convergence of frameability of large systems}
\label{secexponential}

We now show that the 
 probability a configuration of a hypercube is 
frameable approaches one (at least) {\it exponentially fast} in its linear size $L$,
i.e. 

\begin{equation}
\label{exponential}
\mu_{\Lambda,\rho}({\cal{F}})\geq 1-C\exp\left(-\frac{L}{\Xi^u}\right)
\end{equation}
for $L\geq \Xi^u(\rho)$.

In addition to its direct relevance for the square lattice model, this result will be needed for analyzing other KA models. It is instructive to consider
 two ways of proving the exponential convergence.

\begin{description}
\item(i)
The first is via a percolation-like argument, in the same spirit as that used
in reference \cite{CerfManzo}. Consider initially a square lattice $\Lambda$
of linear size $L$, divide it into a set of smaller squares of
linear sizes $l$ and focus on one of them, $\Lambda_\ell$.
%Consider the local environment  of one of these small squares, $\Lambda_l$
%which might not itself be frameable. 
If the four neighboring squares of $\Lambda_l$ 
are framed, then the $3l\times 3l$ square that includes all these small squares can also be framed (see figure \ref{fra6}); other combinations of the nine subsquares also suffice to make the larger square frameable.

\begin{figure}[bt]
\centerline{   
\includegraphics[width=\columnwidth]{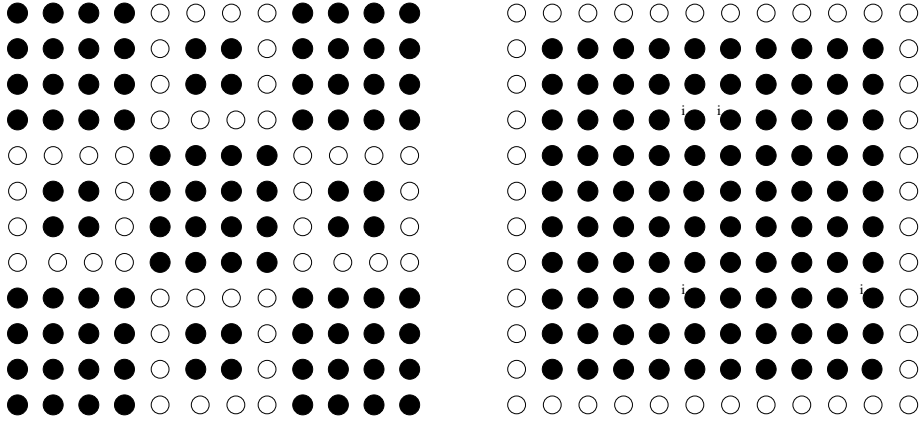}}
\caption{Construction of a large framed square from a partial set of framed subsquares: the two configurations can be connected by a sequence of moves similar to those of  figure \ref{fra2}: starting from the corners that connect two empty edges one can move lines of vacancies up, down, right and left to reach the final configuration. 
\label{fra6}}
\end{figure}

In this way one can grow large frameable squares. The crucial observation is that if the (possibly) unframeable subsquares do not span across the larger square, then  the latter can definitely
be framed. 
%ric.
This condition is simply  the {\it lack} of percolation of unframeable squares. Thus

\begin{equation}\label{in1}
1-\mu_{\Lambda_L,\rho}({\cal{F}})\le P_{perc}(L/l,\mu_{\Lambda_l,\rho}({\cal{F}}),2)
\end{equation}
where $P_{perc} (l,r,2)$ the probability of conventional site percolation
for a square lattice of linear size $l$, with 
occupation probability $1-r$. 
From site-percolation estimates \cite{CerfManzo} 

\begin{equation}
\label{percolation}
P_{perc} (l,r,d)\le \sum_{j=l}^{\infty } 4(1-r)^{j}4^{j-1}=\frac{4}{3}\frac{(3 (1-r))^{l-1}}{1-3 (1-r)}
\end{equation}
Therefore, if for the sub-squares

\begin{equation}
\label{lstar}
\mu_{\Lambda_l,\rho}({\cal{F}})>1-\frac{1}{3e}
\end{equation}

 equations (\ref{in1}) and (\ref{percolation}) 
imply that for the large square

\begin{equation}
\label{expconv}
\mu_{\Lambda,\rho}({\cal{F}})\geq 1-Ce^{-L/l}
\end{equation}
with a positive constant $C$.
Since (\ref{lstar}) holds for $l=\Xi$ --- by the definition of the crossover length --- the desired exponential bound, (\ref{exponential}), is proven. 

\item(ii)
The second method is via  a renormalization group (RG) approach (see
\cite{Reiter} for a similar argument in the context of bootstrap percolation).
Consider a square lattice $\L_L$ of linear size $L$ and divide it into 
 four squares of linear size $L/2$.
If at least three  of the four subsquares  are frameable, then
 the large square is also frameable.  This implies an iterative inequality:

\begin{equation}
\label{iterineq}
\mu_{\Lambda_{2L},\rho}({\cal{F}})\geq \left(\mu_{\Lambda_{L},\rho}({\cal{F}})\right)^4+4\left(\mu_{\Lambda_{L},\rho}({\cal{F}})\right)^3
 (1-\mu_{\Lambda_{L},\rho}({\cal{F}}))
\end{equation}
If 
\begin{equation}
\label{condition}
1-\mu_{\L_{L},\rho}({\cal{F}})\leq \frac{1}{10e}
\end{equation}
then on the next scale up

\begin{equation}\label{iterineq2}
\mu_{\L_{2L},\rho}({\cal{F}})\geq 1-10(1-\mu_{\L_{L},\rho}({\cal{F}}))^{2}
\end{equation}
which implies that condition (\ref{condition}) holds also for $2L$. 
   We can thus iterate
 the inequality (\ref{iterineq2}) to obtain

\begin{equation}\label{iternieq3}
\mu_{\L_{2^nL},\rho}({\cal{F}})\geq 1-1/10\left(10(1-\mu_{\L_{L},\rho}({\cal{F}}))\right)^{2^n}\geq 1-\frac{1}{10}e^{-2^n}
\end{equation}
Since for $L>\Xi(\rho)$ condition (\ref{condition}) holds,  equation (\ref{iternieq3}) gives the desired bound (\ref{expconv}) for large square lattices. \footnote{Note that we have only considered  $L=2^{n}\Xi_{2,1}(\rho)$ for integer $n$, but the results can be readily extended to all $L$.}\end{description}

%These proofs can be easily generalized to the other cases (higher dimensions and values of $s$) in order to show that the 
%that the probability that a configuration of a hypercube is frameable approaches one always (at least) {\it exponentially fast} 
%in its linear size $L$. 

\subsection{\bf Irreducibility for d=3, s=2}
\label{d3}

We now turn to the double-vacancy-assisted cubic lattice case, $d=3$ $s=2$, that was originally studied by Kob and Andersen, and  prove that configurations are almost surely in the same large  irreducible component in the infinite system  limit for any density.
This case can be analyzed by the general discussion of section \ref{ds}, however the study of this more highly constrained three-dimensional case is useful as it introduces,  in a readily visualized situation, the technique we use to extend ergodicity results from smaller to larger values of $d$ and $s$.

Consider a cubic lattice $\Lambda\in \bZ^3$ with size
$L^3$.  Define {\it framed} configurations as those  with all
 the boundary sites empty and define {\it frameable} as those  that, by an allowed 
sequence of elementary moves, can reach a {\it framed} configuration.
Again, frameable configurations with the same number of particles belong to
the same irreducible component.
This can be checked by noticing that the bottom and top planes of vacancies can be
raised and lowered performing the same ``sandwich procedure'' as for the 
 $d=2$, $s=1$ case, but here using empty planes instead of empty rows.

A lower bound on the probability that a configuration is frameable,  $\mu_{\Lambda,\rho}({\cal{F}})$, can be obtained by the following expansion argument.
Consider a  cube of linear size four, which has at its center an empty cube of linear size two. If adjacent to each face of 
the internal cube there is a four by  four square that is frameable  in the  $d=2$, $s=1$, --- single-vacancy-assisted planar --- sense, the whole cube of size four is frameable in the desired three-dimensional double-vacancy-assisted sense. This procedure can be iterated to grow an $L$ by $L$ by $L$ frameable configuration from an empty cube of size two, by requiring that in each subsequent shell all six of the {\it square faces are frameable}.
\footnote{Actually we should require a little more than that all the square faces be frameable:  one needs 
two frameable square of size $L$ by $L$, two of size $L$ by $ L+2$ and two of size $L+2$
 by $L+2$.
We skip these details whose effects are subdominant compared to those we compute.}. 

Letting $\mu_{\Lambda,\rho}({\cal{F}}^0)$ be  the probability of frameable configurations grown from a nucleus at the origin, a series of lower bounds for $\mu_{\Lambda,\rho}({\cal{F}})$ follows beginning from a completely empty $X^3$ nucleus for general $X$
\begin{equation}
\mu_{\Lambda,\rho}({\cal{F}})\geq \mu_{\Lambda,\rho}({\cal{F}}^0)\ge(1-\rho )^ {{X}^3} \prod _{l=X/2}^ {L/2-1} \left(\mu^{2,1}_{\Lambda_{2l},\rho}({\cal{F}})\right)^{6}~~~~
~~\forall~ X
\label{fr3}
\end{equation}
with $\Lambda_{2L}$ is a $2L$ by $2L$ square corresponding to one of the expansion faces. 
From the previous subsection we know that the frameable probability of a square, $\mu^{2,1}_{\Lambda,\rho}({\cal{F}})$,
 is exponentially close to unity  for $L\gg \Xi_{2,1}(\rho)$.  As in the case of bootstrap percolation
\cite{VanEnter} choosing $X>>\Xi_{2,1}(\rho)$
in (\ref{fr3}) makes the pre-factor dominate over the product and we conclude  that 

\be 
\label{prob2}
\mu_{\infty,\rho}({\cal{F}}^0) \sim (1-\rho)^{\Xi_{2,1}(\rho)^3}
\ee
 for $L\gg\Xi_{2,1}(\rho)$; this yields 
 \begin{equation} 
 \xi_{3,2}(\rho) \sim \Xi_{2,1}(\rho)
 \end{equation}
 as the characteristic length scale above which the frameability conditions are met with high probability.

By considering the $O\left(L^3/\xi^3\right)$ possible initial nuclei in a large $L^3$ cube, we can show, by analogy with the two-dimensional case, that 
 $\lim_{L\to\infty}\mu_{\Lambda,\rho}({\cal{F}})=1$ thereby establishing an upper bound for the crossover length $\Xi_{3,2}(\rho)$ above which the large irreducible component dominates:
\be 
\label{prob3}
\Xi^u_{3,2}(\rho)\equiv\xi_{3,2}(\rho)(1-\rho)^{-\Xi^u_{2,1}(\rho)^3/3} \ .
\ee

Up to subdominant corrections,  this means
\begin{equation}
\log \Xi^u_{3,2}(\rho) \sim [\Xi^u_{2,1}(\rho)]^3
\end{equation}

\subsection{\bf Irreducibility for all $d$, $s=1$}
%Single-vacancy-assisted models in general dimensions}
\label{s1}

We now extend the results derived  in the previous two subsections.  First we consider $s=1$ on general hypercubic lattices.

As above, define  {\it framed}
configurations to be those with all their one-dimensional hyperedges
empty and define as {\it frameable} those reachable from  framed configurations.
In the three-dimensional cubic case, for example, the relevant edges are  the twelve edges of a cube.
%\begin{eqnarray}
%&&x_{1}=1,x_{2}=1;x_{1}=L,x_{2}=1;x_{1}=1,x_{2}=L;x_{1}=L,x_{2}=L;\nonumber\\ 
%&&x_{1}=1,x_{3}=1;x_{1}=L,x_{3}=1;x_{1}=1,x_{3}=L;x_{1}=L,x_{3}=L;\nonumber\\
%&&x_{2}=1,x_{3}=1;x_{2}=L,x_{3}=1;x_{2}=1,x_{3}=L;x_{2}=L,x_{3}=L \nonumber
%\end{eqnarray}
As in the cases already discussed, all framed configurations --- and therefore also the corresponding frameable
ones --- belong to the same irreducible component.
This can be seen, for example, in the cubic case, by considering permuting the
occupation variables of two neighboring sites $\{i,j\}$ which lie in a plane parallel  to the ``bottom" plane.
By starting from the corners, the bottom  frame can be raised till it frames the plane that contains the pair $\{i,j\}$.
 Then, after applying on this plane the
 same sandwich technique used for single-vacancy assisted motion on a square lattice,  the desired 
permutation can be performed and the frame then reexpanded and returned to the bottom plane.

The
 growing technique that enables the construction of  larger frameable configurations starting from an empty nucleus, can be generalized via the following observation. Given a frameable hypercube of linear size $L$, to expand it to size $L+2$, $2^{d-1}$ vacancies are needed in the subsequent shell for each of the $2^d$ hyperfaces.
Therefore,
 the probability that a configuration 
is frameable, $\mu_{\Lambda,\rho}^{(d)}({\cal {F}})$,
  can be bounded from below by $\mu_{\Lambda,\rho}^{(d)}({\cal {F}}^0)$, with
\begin{equation}\label{UBS1D}
\mu_{\Lambda,\rho}^{(d)}({\cal {F}}^0)=(1-\rho
 )^{2d}\prod_{{l}=2}^{\infty }
 \left(1-\sum_{j=0}^{2^{d-1}-1}\frac{l^{d-1}!}{( l^{d-1}-j)!j!}
 \rho^{l^{d-1}-j} (1-\rho )^{j}\right)^{2^{d}}
\end{equation}
where the product is over even $l$. 
In the limit $\rho\to 1$, $L\to\infty$ 
\begin{equation}\label{UBS21D}
\mu_{\L,\rho}^{(d)}({\cal {F}}^0)\sim \exp \left(-\frac{c (d)}{(1-\rho )^{\frac{1}{d-1}}} \right)
\end{equation}
where
\begin{equation}\label{UB3S1D}
c (d)= -2^{d-1} (d-1)
\int_{0}^{1}\frac{dy}{y (\log y)^{\frac{d-2}{d-1}}}\log
\left(1-y\left(\sum_{j=0}^{2^{d-1}-1}\frac{(-\log y)^{j}}{j!} \right) \right) 
\end{equation}

As for the square case, there exists a length 
\be
\xi(\rho,d,1)\approx\left(-\log(1-\rho)/(1-\rho)\right)^{1/ (d-1)}
\ee
 such that $\mu_{\Lambda,\rho}^{(d)}({\cal {F}}^0)$ varies little with $L$ for $L\gg\xi$.
Therefore, by considering the $O\left(L^d/\xi^d\right)$ possible distinct positions for the nucleus, an upper bound for the cross-over length follows
\begin{equation}
\label{CLS1}
\Xi_{d,1}^u(\rho)\sim \exp \left(\frac{c (d)}{d(1-\rho )^{\frac{1}{d-1}}} \right)
\end{equation}
As explained at the end of section \ref{d2}, requiring that the number of possible positions for framing nuclei times the probability that  the configuration is frameable about a given  nucleus  is of order one,  is not enough to conclude that 
$\mu_{\Lambda,\rho}({\cal{F}})=1$.
Again, the conditions that a configuration is frameable about two different nuclei, are not independent events. However, provided the nuclei  are sufficiently far apart, these events are almost independent
 since the frameability requirements become weaker and weaker for larger
 sizes. Indeed the requirement to have at least $2^d-1$ vacancies on each of the $2^d$ faces of a frameable hypercube only amounts to $2^d-1$ vacancies among $L^{d-1}$ sites and thus, at fixed density, this becomes less and less restrictive at larger sizes.

We outline an iterative argument by which the desired previous result 
can be proved (see \cite{Reiter} for a similar argument for bootstrap
percolation). Divide the lattice into  $(L/l_{2})^{d}$ hypercubes of linear size $l_{2}$. At
the center of each of these hypercubes consider a smaller hypercube $\Lambda_1$ of
size $l_{1}=l_2/2$.
The probability that the lattice is frameable can be bounded using the
probability that the lattice can be made frameable starting from one
of the $\left(L/l_{2}\right)^{d}$ hypercubes  of linear size $l_{1}$. 
We thus have a bound $\mu_{\L,\rho}({\cal{F}})\geq P_1P_2P_3$, in terms of the probabilities of three events:
 $P_{1}$, the probability that at least one of the
hypercubes of size $l_{1}$ is frameable; $P_{2}$, the probability that this
frameable hypercube can be expanded until the size $l_{2}$ (requiring
$2^{d-1}$ vacancies for each shell from size $l_{1}$ to size $l_{2}$); and  $P_{3}$,
the probability that the frameable hypercube of size $l_{2}$ 
can be expanded until the size $L$ requiring$2^{d-1}$ vacancies
on each shell from size $l_{2}$ to $L$ {\it excluding} the regions of
space occupied by the other small hypercubes of size $l_{1}$. 
The following results can be readily derived:
\begin{eqnarray}
P_{1}&=& 1- (1-\mu_{\Lambda_1,\rho}({\cal {F}}^0))^{\frac{L^{d}}{l_{2}^{d}}}\simeq 1-\exp
{\left(-\mu_{\Lambda_1,\rho}({\cal {F}}^0) {\frac{L^d}{l_{2}^d}}\right)}\nonumber \\
P_{2}&=&\prod_{l=l_{1}}^{l_{2}}
 \left(1-\sum_{i=0}^{2^{d-1}-1}\left(\genfrac{}{}{0pt}{1}{l^{d-1} }{i}\right)
 \rho^{l^{d-1}-i} (1-\rho )^{i}\right)^{2^{d}}\nonumber\\
P_{3}&=& \prod_{l=l_{2}}^{L}
 \left(1-\sum_{i=0}^{2^{d-1}-1}
\left(\genfrac{}{}{0pt}{1}{(l/2)^{d-1}}{i}\right) \rho^{(l/2)^{d-1}-i} (1-\rho )^{i}\right)^{2^{d}}\nonumber\\
&&
\end{eqnarray}
By choosing $l_1\approx\xi(\rho,1,d)$, these give
\begin{eqnarray}
P_1&\approx&1-\exp
\left(-\exp \left(-\frac{c (d)}{(1-\rho )^{\frac{1}{d-1}}} \right){\frac{L^d}{l_{2}^d}}\right)\nonumber\\
P_2&\approx& 1\nonumber\\
P_3 &\approx& 1
\end{eqnarray}
so that $\mu_{\Lambda,\rho}({\cal {F}})$ is of order unity as long as $\exp(-c (d)/(1-\rho )^{1/(d-1)})\left(L/l_{2}\right)^d\gg1$, i.e. for $L\gg\Xi^u_{d,1}(\rho)$ with $\Xi^u_{d,1}(\rho)$ given by Eq.(\ref{CLS1}).
 
\subsection{\bf Irreducibility for all $d$ with $s<d$}
\label{ds}

We now
consider KA models with general parameter $0<s<d$ on a hypercubic lattices  of dimension $d$. These are the non-trivial cases.
[As mentioned earlier, or $s\geq d$, any completely occupied d-dimensional hypercubes are frozen so that these KA models on hypercubic lattices are non-ergodic for any density $\rho>0$. At the opposite extreme, $s=0$ is the normal lattice gas which is trivially ergodic.]

The $s=1$ case was considered in the previous subsection. We now show how the general case can be analyzed, and irreducibility proved, via  a  generalization of the iterative procedure used in section \ref{d3} to extend the $d=2$, $s=1$ results to $d=3$, $s=2$.

Define {\it framed} configurations of $d$-dimensional hypercubes as those with all hyperedges of dimension $s$ empty. [More formally the configurations with no particles on the hyperplanes:
\begin{equation}
\{x_{E_{1}}=1,L\}\times\{x_{E_{2}}=1,L\}\times\dots \{x_{E_{d-s}}=1,L\}
\end{equation}
where $E_{1},\dots E_{d-s}$ for all sets of $d-s$ of the indices 
$1,\dots ,d$.]
%For example the $s=2,d=3$ case corresponds to the planes:
%$x_{1}=1;x_{1}=L;x_{2}=1;x_{2}=L;x_{3}=1;x_{3}=L$.
Inside a frameable hypercube, any move of a particle to a neighboring empty site  
can be achieved by the generalization of the sandwich technique
discussed previously.

Consider a frameable hypercube of size $l$.
If adjacent to each hyperface there is a $d-1$-dimensional hypercube that is
frameable in the $d-1$, $s-1$ sense, then the enclosing d--dimensional hypercube of size $l+2$ is frameable.
Therefore, to bound from below the frameable probability,
$\mu_{\L,\rho}({\cal{F}}^{(d,s)})$, we need to estimate the probability  
that a hypercube of linear size $X$ centered at the origin is empty and that  expanding from it
any subsequent shell is frameable in the $d-1$, $s-1$ sense. This yields
\begin{equation}
\label{kk}
 \mu_{\L^d,\rho}({\cal{F}}^0_{(d,s)})\geq(1-\rho)^{X^{d}} \prod_{\ell=X+1}^{L}
\mu_{\Lambda_\ell,\rho}({\cal {F}}_{(d-1,s-1)})
\end{equation}
where $\Lambda_\ell^{d-1}$ is a $d-1$-dimensional lattice of linear size $\ell$.
The argument given is section \ref{secexponential} for $d=2,s=1$ can be
extended to  the get that the convergence of the frameable probability to one 
is exponential in the linear size of the system and determined by
$\Xi^u_{d-1,s-1}(\rho)$\footnote{Note that the argument for exponential
  convergence requires  the a priori knowledge of convergence to unity for
  the probability of the frameble
component. This has been proven for any $d$, $s=1$ in
  previous section. Therefore (\ref{if}) holds for any $d$, $s=2$ and the
  argument below leads to $\mu_{\infty,\rho}=1$ for these cases. Therefore, one can
  apply again arguments in section \ref{secexponential} and obtain exponential
  convergence for all the cases $s=2$, i.e. (\ref{if}) for $s=3$. And so on iteratively.}

\begin{equation}
\label{if}
\mu_{\Lambda,\rho}({\cal {F}}_{(d-1,s-1)})\geq 1-
C\exp\left[L/\Xi_{d-1,s-1}^u \right]
~~~~\mbox{for}~~ L\gg \Xi^u_{s-1,d-1}(\rho).
\end{equation}
Hence, by choosing $X\gg \Xi^u_{s-1,d-1}(\rho)$, the
product on $\ell$ in (\ref{kk}) is close to unity. 
From generalization of the arguments given for the $s=1$ square lattice case, a non-zero probability of frameability around a given nucleus that converges rapidly to its asymptotic limit for large sizes implies a probability close to unity for frameability of sufficiently large hypercubes around {\it some} nucleus. More precisely that 
$\mu_{\Lambda,\rho}({\cal {F}}_{(d,s)})$ approaches unity in the
thermodynamic limit and is close to one for $L\gg \Xi^u_{d,s}(\rho)$. Since the dominant small factor in (\ref{kk}) is the large power of $1-\rho$, and to make frameability likely one needs a system large enough to contain at least one frameable hypercube of size greater than $\Xi^u_{d-1,s-1}$, an upper bound on the crossover length is 

\begin{equation}\label{CLds}
\Xi^u_{d,s}(\rho)\sim \left[(1-\rho)^{-\Xi^u(\rho,d-1,s-1)^{d}}\right]^\frac 1d \ .
\end{equation}

From the earlier results for $s=1$  in general $d\geq 2$, 
we know $\Xi_{d1}^u(\rho)$: Eq. (\ref{CLS1}).
Therefore, by induction, equations (\ref{CLds}) yield  upper bounds for the crossover length for general $s,\ d$ with  $1<s<d-1$.
The density dependence of this upper bound, $\Xi^u$, is hence

\begin{equation}
\Xi^u(\rho,s,d)=\exp^{\circ s}\frac{C (d,s)}{(1-\rho)^{1/(d-s)}}
\label{xiu}
\end{equation}
 where
$\exp^{\circ s}$ denotes the exponential function iterated $s$ times. Note that the factors of $\ln(1-\rho)$ that appear in \ref{CLds} can be taken into account, as far as the upper bound, by a slight modification of $C(d,s)$.

\subsection{\bf Ergodicity for all $d$ with $s<d$ }
\label{irreducibility}

Thus far, we have only considered the irreducibility of configurations in infinite systems. In this section we prove that the almost-sure existence of an irreducible component  in the thermodynamic limit implies ergodicity (i.e., the third step of the procedure outlined in section \ref{d2}).
We follow the same  strategy as in section  $4$ of \cite{BT}. 
Since the proof is the same for any of the models, 
we temporarily drop the indices $d$ and $s$.
It is first necessary to introduce some mathematical notation and give a rigorous definition of ergodicity for infinite  systems.

Define $\Omega_\Lambda\equiv\{0,1\}^{|\L|}$ as the configuration space of  the
hypercubic lattice  $\Lambda\in\bZ^d$ and configurations $\eta$ as
the elements of $\Omega$, namely sets of occupation numbers $\eta_x\in \left\{0,1\right\}$ for $x\in\Lambda$.
The dynamics of KA models is  a continuous time Markov process with generator $\cL$ acting on local functions  $f:\O_\L \to \bR$ as
\begin{equation}
\cL f \, (\eta)= \sum_{ \genfrac{}{}{0pt}{1}{\{x,y\}\subset\Lambda }{|x,y|=1} } 
\eta_x(1-\eta_y)c_{xy}(\eta) \left[f (\eta^{xy})-f(\eta)\right]
\label{generator}
\end{equation}
with the sum running over nearest neighbor pairs,  $x$, $y$. The quantity in square brackets, 
\be \label{nabla}
\nabla_{xy} f\equiv f(\eta^{xy})-f(\eta) \ , 
\ee 
acts to change the occupation of sites $x$ and $y$
so that the occupation number of the $z$th site in configuration $\eta^{xy}$ is related to the occupations in the configuration $\eta$ by
\be
\eta^{x,y}_z :=
\left\{
\begin{array}{ll}
\eta_y & \textrm{ if \ } z=x\\
\eta_x & \textrm{ if \ } z=y\\
\eta_z  & \textrm{ if \ } z\neq x,y \ ;
\end{array}
\right.
\end{equation}
The jump rates $c_{x,y}(\eta)$ encode the constraints imposed by the KA rules,
namely 

\begin{equation}
\label{KA2} 
c_{xy} (\eta) := 
\left\{
\begin{array}{ll}
1 & \textrm{ if \ }  n_x^{\overline{y}}(\eta)  \leq m \textrm{\ and \ } n_y^{\overline{x}}(\eta) \leq m 
\\
0 & \textrm{ otherwise \ } 
\end{array}
\right.
\end{equation}
where

\be
n_x^{\overline{y}}(\eta) := \sum_{ \genfrac{}{}{0pt}{1}{z\in\Lambda, \ z\neq y
  }{|x-z|=1}}\eta_z  \ \ \ \ ,
\ee
i.e.  $n_x^{\overline{y}}(\eta)$ is the number of occupied neighbors of $x$,
excluding $y$.

The dynamics preserves the number of particles, [i.e. the subspaces with fixed number, $N$, of particles  $\O_{\L,N}:=\{\h\in\O_\L\,\:  \sum_{x\in\L}\h_x=N\}$ are invariant and the  uniform measure $\nu_ {\L,N}$ on fixed $N$ subspaces is invariant].
 However, due to the vanishing of certain rates for $m<2d-1$ ($s\geq 1$), the configuration space, $\O_{\L,N}$, of finite lattices  breaks into
disconnected components: the Markov chain is hence reducible so that the process is not ergodic on these subspaces. On the other hand, for the infinite lattice $\Lambda=\bZ^d$,
the process satisfies detailed balance with respect to the trivial Bernoulli
 product measure, $\mu_{\rho}$, at any density $\rho$. Moreover, from the
 previous section, we know that there exists an irreducible component --- the
 set of  frameable configurations --- that covers the configuration space. However this is not enough to conclude that the system is ergodic. Indeed, while for finite state systems irreducibility of the Markov chain implies that there exists a unique invariant measure and the system is ergodic with respect to it, this is not a priori true in infinite systems. 
For example, Ising models below their critical temperature are
irreducible but not ergodic in zero field because of the coexistence of the two phases each of which lasts for infinite time.   To establish
ergodicity for infinite systems, one needs to prove \cite{Liggett}
that the long time limit of all  
correlation functions --- weighted averages
 over the probability distribution at time $t$, ${\cal P}_t={\cal P}_t\sim\exp
 {\cL t}$ --- approaches those of the Bernoulli product measure for almost all initial conditions, $\eta^I$: more precisely that
$$ 
\lim_{t\to\infty} \: \: 
 \int\!d\mu_\rho(\eta^I)\big[{\cal P}_t f(\eta^I) - \mu_{\rho}(f)\big]^2  =0 
~~~~~\forall f \in L_2(\mu_{\rho})  \ ,
$$ 
where $f\in  L_2(\mu_{\rho})$ if $\int d\mu_\rho(\eta)f^2(\eta)<\infty$
and $\mu_{\rho}(f)$ is the equilibrium average  of function $f$,
namely

\be 
\mu_{\rho}(f)\equiv \int\!d\mu_\rho(\eta) f(\eta) \ ,
\ee
and ${\cal P}_t f(\eta^I)$ is the correlation function $f(\eta)$ averaged over
all possible histories up to time $t$ starting with initial configuration
$\eta^I$.  
%Note that the space of square integrable functions, $L_2(\mu_{\rho})$, includes all functions that only depend on  a finite set of sites. 

By the spectral theorem, convergence  to the equilibrium measure occurs 
if and only if zero is a 
{\it simple} eigenvalue of the generator of the dynamics, $\cL$, i.e. if the only functions, $f_0$, in   $L_2(\mu_{\rho})$ for which $\cL f_0(\eta)=0$ are constant on almost all configurations, i.e. on all except possibly a set of measure zero \cite{Liggett}. 
Note that if the system can be trapped for an infinite time in some  region of
the configuration space, this would not be true:
 the characteristic function of this region would be a non-constant
 eigenvector of $\cL$ with zero  eigenvalue. This corresponds to the natural
 idea that ergodicity breaking is related to the existence of regions of the
 configuration space that are effectively disconnected.  In the Ising model
 case, the function $f(\eta)$ that is equal to unity on configuration $\eta$
 if a majority of the sites have up spins, and equal to zero otherwise, is such an example.

The strategy of the proof of ergodicity of KA models on infinite hypercubic
lattices $\bZ^d$ with parameter $s$ such that $s<d$ 
is the following.
(1) First we prove that, if $f_0$ is eigenvector of $\cL$ with zero eigenvalue,
 $f_0(\eta)=f_0(\eta^{x,y})$ almost surely with respect to $\mu_\rho$ for any pair of sites
$\{x,y\}$, i.e. $\mu_{\rho}([f_0^{xy}-f_0]^2)=0$ where $f_0^{xy}(\eta)\equiv f_0(\eta^{x,y})$. This part of the proof uses as a key ingredient the existence
of an irreducible component which has unit probability in the thermodynamic limit. 
(2) Then we use De Finetti's
theorem \cite{DeFinetti, Georgii} on the decomposition of symmetric measures on product measures to conclude 
 that any such $f_0$ is in fact constant. 

Let us prove (1).
By enumerating frameable configurations, $\eta^1,\ \eta^2,\  \dots \eta^n, \
\dots$, we can rewrite the event that a configuration is frameable as  $
{\cal{F}}:=\bigcup_{n\geq 1}\eta^n$ and define the function $\id_{\cal{F}}$ to
be one if $\eta\in {\cal{F}}$  and zero otherwise.
The result 
$\mu_{\rho}({\cal{F}})=1$ established in previous sections implies

\begin{equation}\label{erg}
\mu_{\rho}([\nabla_{xy}f]^2)
=\mu_{\rho}([\nabla_{xy}f]
^2 \id_{\cal F})
\le \sum_{n=1}
\mu_{\rho}([\nabla_{xy}f]^2 \id_{\eta^n}) ~~~~~~~\forall f\in L_2(\mu_{\rho})
\end{equation}
 where the exchange operator $\nabla_{xy}$ was defined in (\ref{nabla}). From the properties of frameable configurations, for any configuration $\eta\in {\cal{F}}$ one can always 
find a path  $\eta, \ \eta^1, \ \dots, \ \eta^M$ with $\eta^M=\h^{xy}$ that connects $\eta$ to $\eta^{xy}$ through allowed moves, i.e.  with $\eta^{i+1}=\eta^{i;zw}$ for some pair of neighboring sites $\{z,w\}$ and $c_{zw}(\eta^i)=1$.
By telescoping sums and the  Cauchy--Schwartz inequality  \footnote{Note that for any frameable configuration the
  existence of a path that joins $\eta$ and $\eta^{x,y}$ is
  guaranteed. However, its length can be arbitrarily large on some rare
  configurations. Therefore, strictly speaking, one cannot directly apply Cauchy--Schwartz
  inequality which would pull out a non-bounded factor proportional to the
  length of the maximal path. However, by dividing frameable configuration on
sets characterized by the length $\ell$ 
of the minimal path (to perform such move), 
it is possible to bound each term in the r.h.s. of (\ref{erg}) as an infinite
sum (on $\ell$) of terms of
the kind $\ell$ multiplied by terms $\mu_{\rho}(c_{zw}
[\nabla_{zw}f]^2)$. This, together with the observation (see below) 
that each of this
terms must be equal to zero, is sufficient to conclude that the l.h.s. in
(\ref{erg}) is equal to zero.}, 
each term in the  sum (\ref{erg}) can now be bounded from above by a sum of 
elementary exchanges allowed by KA rules, i.e. terms $\mu_{\rho}(c_{zw}
[\nabla_{zw}f]^2)$.
% leading to
%
%\begin{equation}
%\label{giu}
%\mu_{\rho}(\nabla_{xy}f]^2)\leq ~~~~\forall f
%\end{equation}
 Note that we can introduce the jump rate in
the upper bound only since,
as written above, we can choose elementary moves with rate equal to one (never
to zero). This is thanks to the fact that the frameable component has unit
probability, thus we could restrict the mean in (\ref{erg}) as a mean over 
the frameable component, where we know that a path of allowed moves exists for sure.
The hypothesis of sentence (1), namely 
that there is some $f_0$ that is an eigenvector of $\cL$ with zero eigenvalue, 
implies that 
$\mu_{\rho}(f_0\cL f_0)$ is zero.
Moreover, the symmetry $c_{xy}(\eta)=c_{yx}(\eta)$ --- the microscopic reversibility --- implies 

\begin{equation}
\mu_{\rho}(f_0\cL f_0)=-\frac{1}{2}\sum_{\{x,y\}\in\bZ^d } \mu_{\rho}\left(c_{xy}(\eta)\, [\nabla_{xy} f_0]^2\right)\quad .
\end{equation}
Since the right hand side of (\ref{erg}) can be rewritten as a sum of 
such terms, the fact that $\mu_{\rho}(f_0\cL f_0)=0$ implies
$\mu_{\rho}([\nabla_{x,y}f_0]^2)=0$ and completes the proof of (1).

We now turn to the proof of (2).
 Consider the measure $\tilde\mu^{f_0}$ defined
as $\tilde\mu^{f_0}(\eta)\equiv f_0^2(\eta)\mu_{\rho}(\eta)/\mu_{\rho}(f_0^2)$. Since ${\mu_{\rho}}$ is
symmetric under exchanges of particles and vacancies,
i.e. $\mu_{\rho}(\eta)=\mu_{\rho}(\eta^{x,y})$, and $f_0(\eta^{xy})=f_0(\eta)$
on almost all the configuration space, $\tilde\mu^{f_0}$ is also symmetric,
i.e. $\tilde\mu^{f_0}(\eta)=\tilde\mu^{f_0}(\eta^{x,y})$ for any $\eta$ and
any pair $\left\{x,y\right\}$. Therefore, by the De Finetti theorem\cite{DeFinetti, Georgii},  it can be decomposed on product
measures corresponding to different densities. However measures corresponding to different densities are concentrated on different sets of configurations and, 
hence, all the coefficients of the decomposition have to be zero except 
the one corresponding to $\mu_{\rho}$ therefore 
$f_0^2(\eta)/\mu_{\rho}(f_0^2)=1$ which implies that $f_0(\eta)$
is constant \footnote{
Since this procedure can be repeated for any linear combination
of  $f_0(\eta)$ with the constant function, we obtain that the square of any linear 
combination of  $f_0(\eta)$ with the constant function is a constant which implies that
$f_0(\eta)$ is indeed itself constant.} and equal to 
the coefficient of the decomposition corresponding to $\mu_{\rho}$
\footnote{Note
  that for Ising models at low temperature in zero field we cannot
  invoke such a unique decomposition (the hypothesis for De Finetti's 
theorem do not hold since the equilibrium measure is not simply a product) and therefore
$\mu_{\rho}([f_0^{x,y}-f_0]^2)$  does not imply that $f_0$ is constant.}.

\section{Quantitative results}
\label{Quantitative}
In previous sections we have proven that for any parameters $d$ and $s$, with $s<d$,
ergodicity holds in the thermodynamic limit. Moreover, we have determined an upper bound $\Xi^u_{d,s}(\rho)$ for the density dependent
cross over length, 
$\Xi_{d,s}(\rho)$, that separates two different regimes on finite volume
hypercubes $\Lambda\in\bZ^d$ of linear size $L$: if $L\to\infty$ and
$\rho\to 1$ with $L\geq \Xi_{s,d}(\rho)$, than $\mu_{\Lambda,\rho}({\cal{M}})\to 1$; 
if $L\to\infty$ and
$\rho\to 1$ with $L\geq \Xi_{s,d}(\rho)$, than $\mu_{\Lambda,\rho}({\cal{M}})\to 0$, where
${\cal {M}}$ is the maximal irreducible component (i.e. set of configurations
on $\Lambda$ connectable by possible path which has 
larger probability w.r.t. Bernoulli measure at density $\rho$, $\mu_{\Lambda,\rho}$). In other words, for $L\ll \Xi_{s,d}(\rho)$ the
 configuration space is dominated by a single ergodic component, while for
 $L\ll \Xi_{s,d}(\rho)$ different components contribute.

In the next  subsection we find a {\it lower bound} $\Xi^l_{d,s}(\rho)$, which
has  the same form of density dependence as the upper bound.  In the
subsections \ref{optimal},\ref{optimal3} we show that for specific cases, the exact asymptotic behavior of the crossover length can be obtained --- including the numerical coefficients.

\subsection{\bf Bootstrap percolation and lower bounds on crossover length}
\label{bootstrap}

In this subsection we explain how one can obtain a lower bound on the crossover length, $\Xi$, thanks to a relation among KA models
and the well studied problem of {\it bootstrap percolation}.

Consider a random configuration  at density $\rho$ on a d-dimensional
hypercubic lattice $\L$. One can isolate frozen subsets (possibly 
empty) of such a configuration by the following deterministic procedure.
First remove all the particles that have no more than $m$ neighbors, 
then iterate this procedure until no more particles can be removed.
Indeed, as already noticed by Kob and Andersen for the $d=3$ $s=2$ case, 
all the  particles (if any)
that  remain at the end of such 
procedure are frozen: namely they can never move
under KA dynamics starting from the original configuration \footnote{Note that
  the converse is not guaranteed: a priori there could exist sets of particles 
that are thrown away 
during the deterministic procedure, but they are
forever blocked under KA dynamics.}.
This procedure is simply related to conventional bootstrap percolation  
 (\cite{Adler}, \cite{Leath}, \cite{AizenmanLebowitz}) which
corresponds to starting from a configuration at density $\rho$, iteratively 
 {\it adding} particles in empty sites that have fewer than $m$ neighbors, and
 considering any clusters of empty sites that may remain at the end of this
 process. In other words, by exchanging particles and
 vacancies, and hence $\rho$ with $1-\rho$,
one recovers the usual bootstrap procedure.
Let $\mu_{\Lambda,\rho}^{B}$ be the probability that, starting from a random configuration on an hypercube
 $\Lambda\in\bZ^d$ of linear size $L$, a cluster of particles
remains at the end of the modified bootstrap procedure we are
interested in. 
The rigorous results in \cite{CerfManzo} for usual bootstrap percolation 
yield the existence
of a crossover length $\Xi^B (\rho,d,s)\equiv\exp^{\circ s} (K(d,s)/ (1-\rho
)^{d-s})$ such that if one sends simultaneously $\rho\to 1$ and $L\to\infty$
with $L$ growing faster (slower) than $\Xi^B$, $\mu_{\Lambda,\rho}^{B}$ goes
to one (zero).
Furthermore, the convergence to unity in the
regime for $L>\Xi^B (\rho,d,s)$ is exponentially fast.

%is very small for 
%\be \label{bootbound}
%L<\Xi^B (\rho,d,s)\equiv\exp^{\circ s} (K(d,s)/ (1-\rho )^{d-s})
%\ee but it   converges to unity exponentially fast 
To our knowledge only upper and lower bounds on the 
constant $K(d,s)$ are available in the general case, while the exact value has
been recently determined \cite{new-bootstrap} for the case corresponding to our $d=2$, $s=1$.
For the non-trivial models, $s<d$, 
 if clusters of particles remain, they must be  system-spanning clusters (i.e.  clusters that
connect two boundaries of the system): in infinite systems, such frozen clusters must be infinite. 
Since on any finite lattice two configurations with different frozen sets
belong to different irreducible components, the likely presence of 
such system-spanning clusters of frozen particles 
%for $L<\Xi^B (\rho,s,d)$,  implies that for hypercubes smaller than the bootstrap crossover length, the configuration space is broken into many ergodic components with respect to KA dynamics.
%Thus we have 
implies immediately that when $\rho\to 1$ and $L\to \infty$ with $L$
increasing slower
than $\Xi^B(\rho,d,s)$, the probability of the maximal irreducible component 
for KA model ($\mu_{\Lambda,\rho}({\cal{M}})$
is strictly smaller than one. This corresponds to the following 
 lower bound for the KA crossover length:

\begin{equation}
\label{bootbound}
\Xi_{d,s}(\rho)\geq\Xi^l_{d,s}(\rho)=\Xi^B (\rho,d,s) \ .
\end{equation}
Note that the form of the density dependence of this lower bound 
(\ref{bootbound}) is
the {\it same} as that of the upper bound, (\ref{xiu}).
% the difference being
%that the constant $K(d,s)$ is smaller than that in (\ref{xiu}), $C(d,s)$

As mentioned above, for the case $d=2$ $s=1$ recent results have been found in
\cite{new-bootstrap} with a sharp value of the constant:
$K(2,1)=\pi^2/18 $.
On the other hand, from equation (\ref{u2}) we find $C(2,1)= -\int_{0}^{1}
dy\frac{\log (1-y+y\log y)}{y}\simeq 4.4$,
 therefore $K(2,1)<C(2,1)$ and $\Xi^u_{2,1}(\rho)>\Xi^l_{2,1}(\rho)$.
In the next subsection we show how, from a different framing technique, it is possible to determine a stronger upper bound for $\Xi_{2,1}(\rho)$ which turns  out to be equal to the lower bound from bootstrap percolation.

\subsection{\bf Optimal framing and exact crossover length for $d=2, \ s=1$}
\label{optimal}

We now show how exact results for the crossover length of the single-vacancy assisted ($s=1$) KA model on a square lattice can be obtained by an optimal framing construction. 

Define a $W\times H$ rectangle (with $W+H$ even) to be {\it optimally framed} if it
has $\frac{1}{2}(W+H)+1$ vacancies arranged with one in a corner and the
others on alternate
sites of the  
two perpendicular sides that  intersect at that corner, plus any
number of additional vacancies, see figure \ref{fra7}.  {\it Optimally frameable}
configurations are  those that can be
reached from an optimally framed configuration with allowed moves.  
\begin{figure}[bt]
\centerline{   
\includegraphics[width=0.25 \columnwidth]{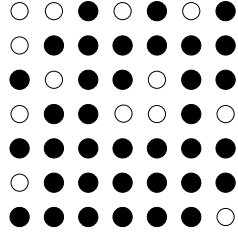}} 
\caption{A $7$ by $7$ optimally framed configuration}
\label{fra7}
\end{figure}

One can show that all optimally framed configurations of a rectangle
with the same particle number belong to the same irreducible
component. Hence all the optimally frameable configurations belong to
the same irreducible component.
This follows by showing that  within an optimally frameable rectangle any nearest neighbor pair of particles can be exchanged and all the 
other particles returned to their starting positions. The sequence of moves to perform a generic such exchange can be constructed by considering the 
 basic moves in figure \ref{fra8}, 
 which allow one to move the nucleus of three vacancies along the row containing alternating vacancies, and those in figure \ref{fra9} which enable one to lower and raise the row of alternating vacancies through the lattice.  
 
 It is straightforward to check that with fewer than $1+\frac{1}{2}(W+H)$ vacancies, large scale rearrangements of the particles in a rectangle are not possible: this suggests that this framing is, at least in some senses, optimal. 

\begin{figure}[bt]
\centerline{    
\includegraphics[width=\columnwidth]{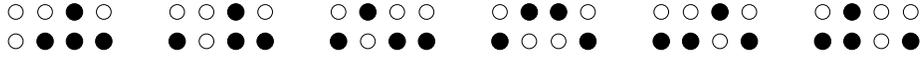}}
\caption{Basic moves by which the three vacancy corner can move along the alternating-vacancy row of an optimally frameable rectangle.
\label{fra8}}
\end{figure}
\begin{figure}[bt]
\centerline{    
\includegraphics[width=\columnwidth]{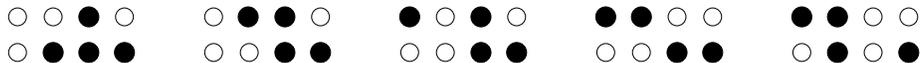}}
\caption{Basic moves by which the row of alternating vacancies in an optimally frameable rectangle can be lowered.
\label{fra9}}
\end{figure}

Following the same strategy as for the simple framing used previously, we now show how larger optimally frameable rectangles can be constructed.
Consider an optimally 
framed $W\times H$ rectangular region embedded in a larger system. The following statements can be checked by direct inspection: if there is a vacancy in a line next-nearest neighbor to one of the rectangular edges parallel to either of the directions $x$ or $y$, the rectangle can be expanded to a $W\times (H+2)$ or $(W+2)\times H$, respectively, optimally framed rectangle; if there is a vacancy in the line segment next to one of
 the rectangle's edges it is likewise {\it expandable} into a $W\times (H+1)$ or $(W+1)\times H$ 
framed rectangle; and
if there is a vacancy next nearest neighbor along a {\it diagonal} from one of its corners,  the rectangle is expandable to a $(W+1)\times (H+1)$ optimally framed rectangle.

Starting from a nucleus of three vacancies in a
two-by-two square, the described expansion procedure can be iterated to grow larger frameable rectangles as long as the
needed vacancies are present at each step. We must now estimate the probability that all the needed vacancies to construct an infinite frameable region centered on a chosen site are present.
%It is straightforward to show that $\ln\mu_{\infty,\rho}({\cal{OF}})$ is dominated by length scales of order $1/v$.
Define $Q^\ell_{k}$ as the probability that, given a $k-2 \times l$ optimally frameable rectangle, it can be expanded to the right to an optimally frameable $k\times l$ rectangle which  now includes the $(k-1)$th and $k$th columns.
From the above observations, we obtain the
following recursion relations

\begin{equation}
\label{Q}
Q^\ell_{k+2}=Q^\ell_{k+1}(1-e^{-v\ell})+Q^\ell_ke^{-v\ell}(1-e^{-v\ell})
\end{equation}
where we have approximated, in the high density limit of interest,
$\rho^\ell\simeq e^{-\ell(1-\rho)}$ (and used the notation
$v=1-\rho $). Defining the ratios of successive $Q$s by 

\begin{equation}
R^\ell_k\equiv\frac{Q^\ell_{k+1}}{Q^\ell_k}
\end{equation} 
we obtain

\begin{equation}
R^\ell_{k+1}=(1-e^{-v\ell})+\frac{e^{-v\ell}(1-e^{-v\ell})}{R^\ell_k}  \ .
\label{R-recurs}
\end{equation} 
%Since the least unlikely way the configuration to expand 
% vedi transfer matrix
%isroughly isotropically,  at any scale $k\approx \ell$.
Since the least unlikely way for the rectangle to expand is roughly isotropically --- as can be shown --- , we expect $k\approx l$ at large scales . Moreover, for $k$ and $l$ large the ratios of probabilities for successive expansions
along the same axis will vary slowly, therefore a natural --- and checkable --- ansatz is 

\begin{equation}
\label{ansatzR}
R^\ell_{k+1}\approx R^\ell_k\approx R(\ell)
\end{equation}
By substituting (\ref{ansatzR}) in (\ref{R-recurs}), we find
%   Thus we can compute the $\ell$ dependence of the ratios by
%taking 
%\begin{equation}
%\label{approx}
%R^\ell_{k+1}\approx R^\ell_k\approx R(\ell) \end{equation} By substituyiting
% (\ref{approx}) into (\ref{R-recurs}) the value of $R(\ell)$ can be easily
%determined

\begin{equation}
R(\ell)\simeq \frac{1-\cal E}{2}+\frac{1}{2}\sqrt{1+2{\cal E} -3{\cal E}^2}
\end{equation}
where

\begin{equation}
{\cal E} =e^{-v\ell} 
\end{equation}
As the rectangle has to be
expanded to infinity in all four directions concurrently, this yields an 
 estimate for the probability $\mu_{\infty,\rho}({\cal{O}})$ that a specific nucleus of three vacancies can be expanded to an infinite optimally frameable rectangle 

\begin{equation}
\mu_{\infty,\rho}({\cal{O}})\sim
\left[\prod_{n=1}^{\infty}R(2n)\right]^4 \sim
\exp\left[2\sum_{\ell=1}^\infty \ln R(\ell)\right] 
\end{equation}
 which, by
replacing the sum over $\ell$ by an integral and changing variables,
yields

\begin{equation}
\mu_{\infty,\rho}({\cal{O}})\simeq\exp{-\frac{2 c_\infty}{1-\rho}}
\end{equation}
with the constant given by 

\begin{equation}
\label{cinf} 
c_\infty=-\int_0^1\frac{d\cal E}{\cal E}
\ln\left[\frac{1-\cal E}{2}+\frac{1}{2}\sqrt{1+2{\cal E} -3{\cal
E}^2}\right]=\frac{\pi^{2}}{18}\cong 0.55 \ .
\end{equation} 
The knowledge of $\mu_{\infty,\rho}({\cal{O}})$ gives, as in previous sections, an 
upper bound for the crossover length 
\be
\Xi_{2,1} (\rho)\leq
\exp{\frac{c_\infty}{(1-\rho)}} \ .
\ee
 Since the constant $c_\infty$ (\ref{cinf})
has exactly the same value as the bootstrap constant $ K(1,2)$
 in \cite{new-bootstrap}, the upper
bound for $\Xi$ obtained by the above framing procedure coincides
with the lower bound provided by bootstrap results, thus giving the {\it
exact} asymptotic high density  behavior of the crossover length, up to subdominant pre-factors.  This proves that the optimal framing procedure
captures the dominant mechanism that restores ergodicity in large systems at high densities; this justifies referring to it  as {\it optimal framing}.

\subsection{\bf Optimal framing for d=3 s=2}
\label{optimal3}

We have shown that the framing procedure devised in
 section \ref{d2} for the two-dimensional case is not optimal: 
indeed 
 it does not capture the dominant ergodicity restoring mechanisms and 
therefore gives a non optimal bound for the crossover length, $\Xi$. 
(see also section \ref{picture}). 
However, we were able to construct a different
framing procedure which does capture the dominant mechanism. This was done through the construction of a
lower-dimensional {\sl basic structure} (lines of alternated
 particles and vacancies) in which each particle  is blocked
 but only barely so: an {\sl additional lower-dimensional vacancy 
cluster} (an extra vacancy in the square lattice case) can be moved
throughout the whole structure.

We now construct a framing that we conjecture is optimal for the $d=3$
$s=2$ case.  Consider a cube of linear size $L$ which is 
completely filled with the
exception of a plane $L\times L$ on
which the configuration is given by the repetition of the unit cell of eight
sites drawn in figure \ref{opt3}. It is possible to check that all the
particles on such a plane can never move. However, if one adds two vacancies
in place of e.g. the first particles on the two rows of one of the cells, then
all the particles on the plane can be moved.
\begin{figure}[bt]
\centerline{
\includegraphics[width=0.2\columnwidth]
       {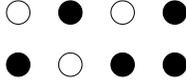}}
\caption{Rectangular unit cell for the basic planar structure for the $d=3$,
  $s=2$ case
\label{opt3}}
\end{figure}
Consider a cube which has three orthogonal faces with the above defined basic planar
structure. We conjecture that with an extra set of $O(L)$ vacancies
associated with the linking edges and $O(1)$ with the corner, the planar
structures can be moved all across the cubic lattice and allow any exchange
inside, i.e. they play the same role as the completely empty faces of
frameable cubes defined in \ref{d3}. 
If this is correct, by requiring $9/8 L^2+ O(L)$ vacancies 
one can construct a configuration for a cube of linear size $L$
 in which all exchanges
are allowed. We conjecture this to be the minimal
requirement, i.e. the above defined framing procedure to be the optimal
one. 

Note that the calculation of the probability for such a  framing procedure to
be expandable should yield an upper bound on the crossover
length which is better than, although  of the same form as,
the one we obtained in
(\ref{prob3}), since fewer vacancies are required in successive layers. However, a rigorous lower bound --- and hence the putative exact value of $C(3,2)$ if the framing described here is indeed optimal ---  is not known; the exact value of $K(3,2)$ for the corresponding bootstrap percolation has not been determined.\footnote{Note that 
knowledge of $C(3,2)$ from framing arguments
would give  a new upper bound on the value of
$K(3,2)$ for the Bootstrap problem and perhaps a hint of its exact value
(recall that $K(2,1)$ coincides with the constant coming from the
correspondent KA problem calculated with the optimal framing).} 
%In principle one could  check test numerically (e.g. by comparing with the
%numerical results in \cite{Pitard}) if the coefficient arising 
%from the
%optimal framing argument is correct although the extrapolation to large sizes is likely to be extremely difficult because of the double exponential dependence of $\Xi$ on $1/(1-\rho)$.  

\section{Physical picture of the cooperative KA dynamics at high density}\label{picture}

In this section, for completeness, we summarize the physical picture that
we derived for the high density dynamics of the KA. This has
already been discussed in \cite{letter} and the derivation of some 
results on the dynamics will be presented in a forthcoming
publication \cite{long}.
We focus for simplicity on the $s=1$ $d=2$ case leaving discussion of 
the natural generalization to any  of the $s<d$ KA models on hypercubic lattices 
to the end. 

As already noted, in the interesting cases there are no finite clusters of vacancies that can
freely move in an otherwise completely filled lattice. 
However, in  a frameable region of size $l=\xi \sim  \ln (1/
(1-\rho ))/ (1-\rho )$, we know that: (1) all particle configurations
inside the region can be reached from each other via an allowed
(although long) path of elementary moves;
(2) a frameable region can move one step in any direction if it can 
expand  one step in that direction (it expands and then leaves a 
vacancy behind on the opposite side when it moves). (3) a frameable region of size $\ell=\xi \sim  \ln (1/
(1-\rho ))/ (1-\rho )$ is likely to find at least one vacancy in each of $\ell$ consecutive line segments along each of the lattice directions. Thus frameable regions of size $\xi$ can 
expand and, hence move,  $\xi$ steps, a distance similar to their diameter. 
%Because of the percolation 
%of such typical regions of size $\xi$, 
%frameable regions of size $ \xi $ can typically move throughout the system.
For this reason, we refer to them as mobile  cores \cite{letter},
although it may be a bit misleading, they have  been called defects in other
contexts. 

The motion of frameable cores is, on long scales,   like simple diffusion in a random
medium. The size of mobile cores, $\ell=\xi \sim  \ln (1/
(1-\rho ))/ (1-\rho )$, is such that the random environment seen by the
cores is the minimal one in which diffusion is possible. For smaller
values of $\ell$ the probability to find the necessary vacancies in the
neighborhood of a core is too low to guarantee that it can diffuse.
In contrast larger frameable regions with $\ell>\xi$ could diffuse,  but these would contribute less to the long-time dynamics because they are both rarer and move more slowly than cores of size $\xi$. [Indeed, larger frameable regions will tend to decompose into one (or more) minimal mobile cores, and leftover vacancies that cannot move without assistance of a mobile core.] \\
%$This naturally leads to the question of what is the typical time scale
%for motion of a core. 
The diffusion coefficient of a tagged particle, $D_S$,  
one of the key physical quantities analyzed in simulations and
experiments, will be given by the spatial density of the cores divided by the
typical time scale for their motion, assuming approximatively  independent core diffusion. 
The density of cores of size $\xi \sim  \ln (1/
(1-\rho ))/ (1-\rho )$, is just the frameable probability around a given origin,
 $\mu_{\Lambda,\rho}({\cal{F}})\sim \Xi^{-d}$. The form of the asymptotic behavior of $\Xi$  in
the high density limit was found for any $s<d$, and, for  the $d=2$ $s=1$ case, the coefficient that enters $\ln \Xi$ was  calculated
exactly see section \ref{Quantitative}.\\
The typical timescale on which cores move has been discussed in \cite{letter}.
Since all frameable configurations are equiprobable, 
this is proportional to $\exp (\Delta S)$, 
 the ratio of the number of accessible (frameable)
configurations of the core to the number in the most severe bottleneck
in the configuration space of the core \cite{letter,long}.
The worst case scenario for a minimally frameable region of size $\ell$ would occur if  the bottleneck corresponded to a {\it single} 
configuration, (the framed one) leading to $\Delta S=S_{\rm total}\simeq \ln
\ell !$  since the total number of frameable configuration is of order $\ell !$. However, as explained in \cite{letter} (and to be
detailed in \cite{long}) the bottleneck
is not nearly as tight: cores do not have to go through 
framed configurations in order to rearrange or to move.
Numerically we have found that the time scale increases more slowly
than exponentially in $\ell$ and its growth is compatible with an analytic  
argument for the entropy bottleneck, which yields at least a lower bound. This result is  obtained by noting that, in order 
to fully equilibrate, an optimally
frameable square of size $\ell$ has to pass through
%configurations in which a minimally frameable square of size of order
%(at least) $1/\va \approx \xi/\ln\xi$ has its nucleus in a {\it
%corner} of the square (line segments of lengths less than $1/\va $, such as the edge of such a square, are typically
%completely filled) .
configurations with the nucleus (the three vacancy element, see
Fig. \ref{fra7} and section \ref{optimal}) in a {\it
corner} of the square. 
Using a transfer matrix technique we obtain for
$\ell\times\ell$ optimally frameable squares an entropy difference
between configurations with the nucleus in the center (which are typical) and those with the nucleus in the
corner asymptotically equal to
$\Delta S \approx \Upsilon\sqrt{\ell} + \alpha \ln \ell + C$ with
$\Upsilon= 2\sqrt{6+\sqrt{22}}-2\sqrt{3}\cong 3.075$ and $\alpha$
computable in principle. \cite{long}  This leads, via the observation that the worst blockages are 
typically when $\ell\sim 1/(1-\rho)$  to a typical timescale 
for movement of optimally frameable regions 
\begin{equation}
\ln\tau_D \sim 1/\sqrt{1-\rho} 
 \end{equation}
 for the $s=1$ square lattice model.
 We conclude that 
the dominant contribution to self diffusion of a tagged particle, $D_{S}$,
arises from the low density of mobile cores rather than the long time needed
 for them to move of the order of their diameter.

The arguments given above can be generalized to any KA models  on hypercubic
lattices with $s<d$. The conclusion is that, to leading order in $(1-\rho )^{- (d-s)}$,
the $s-$iterated logarithm of the self
diffusion coefficient scales as the $s-$iterated logarithm of
$\mu_{\Lambda,\rho}({\cal{F}})$. 
Correspondingly, we expect the typical
relaxation times $\tau$ to scale with one over the density of cores, 
i.e. as the inverse of $D_S$ \footnote{Strictly speaking our results applies directly to $D_S$.
The structural relaxation time could, and very probably will, affected by the slowest 
regions in the systems which instead are not relevant for the self-diffusion. Hence,
we expect a structural relaxation time larger than $1/D_S$, i.e. a decoupling between 
structural relaxation time and self-diffusion as indeed the numerical results 
of Kob and Andersen already suggested originally \cite{KA}.}
.
In particular, thanks to the exact
 result in section \ref{optimal}, for the $d=2$ $s=1$ model we obtain
\begin{equation}
\lim_{\rho \rightarrow 1} (1-\rho )\ln D_{S}=\pi^{2}/9 
\end{equation}
 a prediction which we have successfully verified by 
numerical simulations \cite{letter}. 
On the other hand, for the case $d=3$, $s=2$, the dependence on density for 
different relaxation times $\tau$ have
been considered in \cite{Berthier,Pitard} and our prediction have been
successfully checked,
namely data are well fitted by the scaling $\tau\propto \exp\exp C/(1-\rho)$ 
 with $C$ constant.\\ 
%Numerical results are compatible with the above results for self-diffusion 
%\cite{Berthier,Pitard,letter}. 
Note however that there may be
very substantial  corrections to the leading asymptotic behaviour at obtainable densities.
Indeed, it is known that there are large
corrections to the asymptotic law derived for bootstrap
percolation \cite{Adler}. 
Therefore, it would be
useful to compute sub-leading corrections to all the analytical
expressions derived here. First, one should take into account
the 
 timescale for movement of a core, discussed above. For example in the $d=2$
$s=1$ case corrections to $(1-\rho )\log
D_{S}$ from this will be  large ---- probably 
of order of $\sqrt{1-\rho }$.  
Furthermore, one should take into account corrections to the asymptotic form of the optimal
framing probability: we expect these  
to be negative as  occurs for bootstrap
percolation where the  numerical results for $c_{\infty}$ \cite{Adler} 
correspond 
to roughly one half of the exact result\cite{new-bootstrap}. [Note that recent  progress on
the calculation of such corrections for  bootstrap percolation has been made.] \cite{DeGregorio}).
A rough cancellation between such a negative correction to the density of the mobile cores, and 
the positive one  arising from the typical timescale, 
 might explain
why, in the measured range of densities,  $\ln D_{S}$ seems to be closer to
its asymptotic value than is the crossover scale $\Xi$.  \\

The mechanism that we have identified for the long-time scale dynamics, certainly 
provides a lower bound on $D_{S}$ for $\rho\to 1$ ; this could be probably
be proved along similar lines \cite{long}. But how do we know that there is not a faster mechanism which dominates? There are two reasons to strongly believe that there
are no faster mechanisms: First, we have proven that the finite
size cross-over between irreducible and non irreducible dynamics takes
place exactly when the probability of having at least one core in the
lattice crosses from small to large. This  strongly suggests that a single core is
necessary, as well  as sufficient, to relax the system. This is exactly analogous to 
the behavior for a standard lattice gas in which $D_{S}\propto
(1-\rho )$ at high density and the change from frozen to
unfrozen  takes place when there is at least one vacancy in the region. Secondly,  rigorous
analysis of non-cooperative KA  models\cite{SelfToninelliBiroli} for which finite clusters of $q$ vacancies can move freely, (such as $s=1$ on a triangular lattice, for which $q=2$) show that the self-diffusion coefficient scales as $D_{S}\propto (1-\rho
)^{q}$ in the high density limit. The cooperative cases can be
understood roughly in terms of a density  dependent $q$: at any fixed density there 
are no free moving clusters but there is a characteristic sized 
group of vacancies --- a mobile core --- that can indeed diffuse since it
typically finds enough outer vacancies in each direction (Of course
the environment is random and there are some places in which a core
cannot enter without changing its size. However these regions would not be  
important for the self-diffusion coefficient even if they were forever frozen,
as happens for diffusion in a disconnected random  environment when there is a percolating cluster.). At each density, one needs to have $q \propto \xi (\rho )$,  in order to be sure 
that the diffusion of cores is possible: this yields a mobile core density of  
the form we have found.   

Up to  now we have focused on the behavior at high densities,
in particular  the asymptotic value of relevant quantities in the limit $\rho\to 1$. 
However, as discussed in more detail in \cite{letter, long}, for some lattices, especially with $s>1$,  the asymptotic behaviour
will  not be relevant except on extremely --- perhaps inaccessibly --- long time scales. In practice, much of the 
slowing down as the density increases could occur before this asymptotic
regime. In particular, rapid crossover could occur   near an almost transition
---  sometimes called an {\sl avoided transition} \cite{Kivelson}
 --- near an almost-critical density ( which would depend on the lattice and the value of $s$). 
Such behavior  could  account, for example,
 for the apparent power law  ``vanishing" of the
self diffusion coefficient observed as  $\rho\to \bar\rho\simeq
0.881$  in simulations of the three-dimensional case with $s=2$.

A simple argument suggests that such a cross-over might exist. 

We again first  focus on 
the two-dimensional case with $s=1$. As already mentioned in \cite{letter},
in this case a small (fixed-size) cluster of vacancies can freely diffuse 
on a second-neighbor-percolating clusters of other vacancies. 
Therefore,
when the vacancy density is above the correspondent percolating threshold,
diffusion can occur without substantial cooperativity. But the contribution to the
self diffusion coefficient from this mechanism shrinks to zero as the
vacancy percolation threshold is approached from above (from below, in terms 
of particle density). At higher densities, the mechanism behind diffusion crossesover to 
the cooperative one discussed in this paper.\\
For the three-dimensional case with
$s=2$, from the construction in section \ref{optimal3}, we
 expect that the approximate transition seen in numerics  may well be related to
percolation of surfaces on which the vacancies are nearly 
connected like those on the surfaces of optimally framed cubes.
%this is dual to normal percolation in 3D? 
 
The above percolation  arguments are only qualitative and we do not expect that 
the density at which the percolation transition takes place will correspond
very well with  the density at which the apparent transition takes place.
In next section we present a quantitative analysis of an actual transition in KA models. We show that  in finite
dimensions, apparent transitions with rapid crossovers could arise as  ``ghosts" of actual dynamical transitions that occur on Bethe lattices --- or, equivalently,  within the Bethe approximation for real lattices. This suggests a way to study  ``avoided transitions": by using the Bethe lattice results and perturbing around this 
``mean-field" limit.

\section{\bf KA models on Bethe lattices}
\label{sectionbethe}
Bethe lattices are defined as a infinite  loop-free graphs with 
fixed connectivity $z=k+1$.  The main motivation for studying statistical mechanical models on Bethe lattices is that such ``lattices" are often considered to be a good approximation of more realistic lattices, e.g.  hypercubic,  in the limit of high dimensionality: the Bethe lattice models are then considered to be a type of  
``mean field" --- or more accurately ``uncorrelated field" --- approximation. 
But their tree--like structure enables analytic calculations  via  recursive procedures, 

As we will explain in detail, the behavior of KA models on Bethe lattices is qualitatively different from 
that on hypercubic lattices  obtained in previous sections: on Bethe lattices  an ergodic/non--ergodic
transition takes place at a non-trivial critical density. Nevertheless, exact results on these lattices, when combined with those  for hypercubic
lattices,  should be useful both for better understanding of the latter, and as  background for considering how    mean-field-like  scenarios
for glass transitions might --- or might not --- be extended  to real glasses, or at least to some more realistic models.

\subsection{\bf  Bethe lattices}
\label{bethedefinition}

There are subtleties in the definition of Bethe lattices that are important to note.
Bethe lattices with connectivity $k+1$ are {\it locally} identical to the infinite size limit
of random ``c-regular" graphs with $c=k+1$ \cite{graph} which are uniformly drawn from the set of
graphs with connectivity $c$ for each site with neither multiple
edges (no two edges joining the same pair of sites) nor  loop-edges that join
a site to itself. Since in the 
limit of  large number of sites, with high probability there are no finite loops that go through a
given set of vertices, a typical such random graph looks locally
(i.e.: on any finite length scale) like a Cayley tree  with a fixed branching ratio. [Recall that a Cayley tree with connectivity $k+1$ is
constructed by taking $k+1$ rooted trees with branching ratio $k$ and connecting all the $k+1$ roots to a  new site, which can be considered as
the origin.]
Nevertheless,  for macroscopic properties the 
presence of loops that exist in even very large random graphs is crucial since it induces geometric
frustration and  
ensures a statistically homogeneous
structure which prevents the pathologies that arise for Cayley trees
for which a positive fraction of sites are on the boundary which causes extreme sensitivity to boundary
conditions.  A Bethe lattice corresponds  
to
considering a Cayley tree of size $L$, focusing on a core of size
$\ell$ around the origin and taking the limit $L\to \infty$ before the limit $\ell\to
\infty$, or with the simultaneous limit with both lengths becoming infinite but $\frac{\ell}{L}\to 0$.  Note that this is equivalent for models with local dynamics to considering first the limit $L$ to infinity with random initial conditions, and then considering the long time limit: as such, it is quite physical.\\
The following analysis holds both for random c-regular graphs with $c=k+1$
and for Bethe lattices, up to the subtleties discussed above.

\subsection{\bf Existence of frozen phase }
\label{existence}

KA models on a Bethe lattice with connectivity $k+1$ are defined as in section \ref{definition}, with $0<m<k$ and $s=k-m$.  As usual, it is convenient to  arrange the lattice as a tree with $k$
branches going up from each site and one going down. Before proceeding with a more careful analysis, we first give an argument that on Bethe lattices, in contrast to hypercubic lattices, infinite frozen clusters almost always exist at sufficiently high densities, $\rho$.

Rough analogs of the infinite networks of frozen fully-occupied
slabs that can occur on hypercubic lattices with $s<d$, are fully
occupied infinite subtrees with branching ratio $m=k-s$ and hence
coordination number $m+1$: none of the particles on such a
fully occupied subtree can move. If a Bethe lattice with coordination
number four is arranged with the four bonds coming out of each vertex
forming a cross, then all the vertices of a fully occupied three-sub
tree will appear to have a line going straight through them, and
another half line that either terminates at  the vertex in a T-junction, or also continues through it. Such trees are thus
somewhat similar to a network of bars on a square lattice.  In
another sense, however, because they have no loops, they are more similar to a {\it single} two-wide
infinitely-long solid bar on a square lattice.  Because, for
$m<k$, the number of potential $m$-sub trees is exponentially large
--- in contrast to the number of infinitely long straight
slabs on hypercubic lattices --- it is natural to expect that a fully
occupied  frozen subtree will almost always exist at high
densities. This can be shown by a simple generalization of the
analysis for conventional site percolation on Bethe lattices \cite{betheperc,Leath}. 

Consider a site, call it $i$, 
and the probability, $Q'$, that it is both occupied and is the root of a fully occupied $m-$subtree on the portion of the Bethe lattice above it.  As the existence of such a subtree requires the existence of similar subtrees rooted on at least $m$ of the sites above site $i$, and these are independent events, each with some probability $Q$, we can write a simple recursion relation:
\be
Q'=\rho \sum_{j=k-s}^k Q^j (1-Q)^{k-j} \left(\genfrac{}{}{0pt}{1}{k }{j}\right) \ .
\ee
This has the trivial fixed point solution $Q=0$ for any density, but, above a critical density, $\rho_T$, it has an additional non-trivial fixed point solution, $Q^\ast(\rho)>0$. 
For the simplest interesting case, $k=3$, $s=1$ --- loosely approximating the coordination-four square lattice ---, one can easily solve the fixed point equation to find 
$\rho_T=\frac{8}{9}$, a discontinuous jump at $\rho=\rho_T$ to a non-zero value:
$Q^\ast(\rho_T)=\frac{3}{4}$ and above $\rho_T$ a square-root cusp increase in $Q^\ast(\rho) $ which eventually saturates at $Q^\ast(1)=1$.  Other cases with $m\geq 2$ behave qualitatively similarly, in particular also exhibiting a discontinuous onset of the fraction of sites in the frozen sub-tree --- simply related to $Q^\ast$ ---  at a non-trivial critical point at $\rho=\rho_T(k,m)$.  For models with $m=1$, i.e. $s=k-1$, the existence of an $m$-subtree is the same as the existence of an infinite fully occupied cluster and the behavior of $Q$ is thus identical to that of conventional percolation: there is a critical point at $\rho_T=\frac{1}{k}$ with $Q^\ast$ rising linearly from zero as $\rho$ increases further.

What is the significance of these results for KA models?  As do other percolation arguments, these provide useful bounds: the existence of an infinite fully occupied $m$-tree on a $k$-Bethe lattice, i.e. $Q^\ast>0$,  is a {\it sufficient} condition for the existence of an infinite frozen cluster in the corresponding $k,m$ KA model. Thus we can immediately conclude that a frozen phase is possible in these Bethe lattice models at least for $\rho>\rho_T$. We do not, however, yet have a necessary condition for frozen clusters: as can be seen, these can have isolated vacancies within them  at least at sites with more than $m+1$ frozen subtrees nearby. 

\subsection{Dynamical transition}
\label{bethetransition}
We now analyze the behavior with the actual KA constraints by a generalization of the downwards iterative procedure used above.

Consider a particular site and define the following events:
\begin{description}
\item (i) 
The site is occupied by a particle which can never move up as long as the site below it is occupied; denote by  $Y$ the probability of this event.
\item (ii)
 The site is {\it frozen}: occupied by a particle which can never move up {\it even if} the site below it is empty; denote by $F$ the probability of this event, which is a subset of event (i).
 \item (iii)
 The site is empty but {\it blocked} in such a way that  particles on lower sites can never move up to the site; denote by $B$ the probability of this event. 
\end{description}
By using the tree--like structure and its symmetry under exchange of branches, 
%and neglecting loops
 one can write iterative equations for the (primed) probabilities for the sites in one layer, in terms of the (unprimed) probabilities for the sites in the layer immediately above:
% valgono  in the thermodynamic limit $N\to\infty$
% going down in the tree and look for fixed point of these equations.
%The fixed point (I drop the apice $^*$ by indicating fixed point simply with $Y,B,F$) should satisfy

\begin{eqnarray}
\label{p123}
Y'&=&\rho \left(\sum_{j=k-s}^{k}Y^{j}(1-Y)^{k-j}\left(\genfrac{}{}{0pt}{1}{k }{j}\right)
+\sum_{j=s+1}^{k}B^{j}Y^{k-j}\left(\genfrac{}{}{0pt}{1}{k }{j}\right) \right)\nonumber\\
B'&=&(1-\rho)\sum_{j=k-s+1}^{k}F^{j}(1-F)^{k-j}\left(\genfrac{}{}{0pt}{1}{k }{j}\right)
\nonumber\\
F'&=&\rho\left(\sum_{j=k-s+1}^{k}Y^{j}(1-Y)^{k-j}\left(\genfrac{}{}{0pt}{1}{k }{j}\right)
+\sum_{j=s}^{k}B^{j}Y^{k-j}\left(\genfrac{}{}{0pt}{1}{k }{j}\right)\right) \ .
\end{eqnarray}
These can be alternately written for $Y$ in one layer as a function of the three $Y$s in the three next higher layers.

We are particularly interested in stable fixed points of the iterative equations which correspond to the desired  behavior in the interior of the system; the fixed point equations can be reduced to independent 
 polynomial equations, with $\rho$ dependent coefficients, for any one of the $Y$, $B$, or $F$.  For ease of notation, except when it might be confusing, we will denote fixed point probabilities simply by $Y,\  F, \ B$ and $Q$. 

We have earlier shown that there exists a frozen phase at high densities: in particular for $\rho>\rho_T$ the probability $Q$ is non-zero. Since when the event corresponding to $Q$ occurs, the particle is definitely frozen, we know that $F\geq Q$; $F$ non-zero then implies from (\ref{p123}) that both $Y$ and $B$ are also non-zero at high densities.   

To show that all the restricted events have zero probability at low densities, it is sufficient to note that $Y',\ F'$ and $B'$ all depend at most linearly on $Y,\  F$, and $B$ (as the other powers in the polynomials are greater than unity), and two of  the coefficients are $\rho$. Thus at  
sufficiently low densities,  $Y$ must decrease on iteration by a factor that is no larger than a multiple of $\rho$.  Thus the only fixed point at low densities is $Y=B=F=0$.

We have thereby shown that a dynamical phase transition must exist for Bethe lattice KA models with $0<s<k$. We now show that, except for $s=k-1$, this transition is discontinuous.

\subsubsection{\bf Discontinuity at transition for $s<k-1$}
\label{discontinuity}

For KA models with 
$s<k-1$, i.e.:  $m\geq 2$, one can show straightforwardly that the transition is discontinuous. Focus on one site, call it $x$. A particle on $x$ can move up even if the site below it is occupied, if at most one of
the  $k$ sites above $x$ and the $k^2$ sites  one level further up  are occupied. 
In such a situation, the marginal event whose probability is $Y$ does not occur.  For the empty sites in the two levels above $x$, the marginal occupied event does not occur either and the probability of such sites is thus at least $1-Y$.   With all probabilities having their fixed point values, these observations imply that
\begin{equation}
\label{ineq}
 Y\leq 1-(1-Y)^{k+k^2}-(k+k^2)Y(1-Y)^{k+k^2-1} \ .
\end{equation}
But this inequality cannot be obtained if   $Y$ is positive but small, since for $0<Y\ll 1$ it would become essentially $1\leq (k+k^2-1)(k+k^2)Y$, which cannot hold for small  $Y$. Therefore, the fixed point value of $Y$ cannot go continuously to zero: it must jump to a non-zero value at  the critical density, $\rho_c$.

Note that this argument does not apply for the case $s=k-1$. This is because,
even if all but one of  the sites immediately above $x$  are empty, a
particle on $x$ can still not move up if the site below it is occupied (the inequality (\ref{ineq})  in this case does not have the second term on the right hand side; this invalidates the rest of the argument).

\subsubsection{\bf Continuity of transition for s=k-1}
\label{continuity}

Now consider the most highly constrained non-trivial case:  $s=k-1$. In contrast to the less constrained models,  for a particle to be frozen in this case, it need only belong to an infinite cluster of particles with two or more neighbors, i.e. to a percolating cluster in the  conventional sense of nearest neighbor percolation. This means that for $m=1$, the critical density for the KA model should be  the same as for percolation, $\rho_c=1/ k$ \cite{betheperc}, and the transition will be continuous with $Y$ growing linearly just above $\rho_c$. By solving equations (\ref{p123}) for the simplest such cases, $k=2$ and $k=3$, this  can be  checked directly. More generally, for any $k>2$ with $s=k-1$, to quadratic order when it is small, $Y'$ depends only on $Y$ and can be seen to attain a fixed point value $Y\approx \frac{2}{k-1}(\rho-\rho_c)$ just above the transition at $\rho_c=1/k$; this non-zero $Y$ induces  non-zero, but quadratically small, values for both $B$ and $F$.

\subsubsection{\bf Connection to bootstrap percolation}
\label{betheboot}

As for hypercubic lattices, bootstrap percolation yields bounds for the behavior on Bethe lattices.
Before analyzing in detail the character of the dynamical transition,
we illustrate this connection  which can elucidate
 the physical mechanism underlying the dynamical transition of KA models. 

The relevant bootstrap percolation process is similar to that given
earlier for hypercubic lattices (and introduced by Kob and
Andersen \cite{KA}): From a  random initial configuration with density $\rho$,  remove all the particles that can move according to KA rules and iterate the process until no more particles can be removed. If some particles survive at the end of this procedure, they must be frozen in their initial configuration under the  KA dynamics.
Define  $p^{B}$ as the probability that at the end of the procedure
an infinite particle cluster remains. 
By the above observation, if $p^{B}>0$ the
 KA model at density $\rho$ is not ergodic. 
Let $\rho_c^{B}$ be the critical density, if any, at which a {\it bootstrap percolation transition} takes place, namely $p^{B}=0$ 
 for $\rho<\rho_c^{B}$ and $p^{B}>0$  for $\rho>\rho_c^{B}$.
If there exists such a bootstrap transition, a dynamical transition for the corresponding KA model must also occur and the corresponding critical density satisfies $\rho_c\leq \rho_{B}$. Moreover, a reasonable Ansatz, is that 
the two transitions and therefore the corresponding critical densities, in fact,  
 {\it coincide}.
In the rest of this section we present an argument supporting this conjecture.

Let $x$ be a site on which a particle is frozen  with respect to KA dynamics, i.e. a particle that cannot  move on any time scale after the thermodynamic limit has been  taken.
Than $x$ must have {\it either}: more 
than $m=z-s-1$ neighbors that are occupied and  frozen,  or, if $m$ or  fewer of the neighbors are frozen (and thus the particle on $x$ could potentially move), {\it all} its unfrozen neighboring sites, either empty or occupied, must  themselves  have more than $m$ occupied frozen neighbors. If this were not the case,  
 one could move all the unfrozen neighbors 
% or next-to-nearest neighbors 
out of the way, and then move the putatively frozen particle from site $x$.
  Moving the neighbors at the same time  is possible in such a situation because
 the constraints on them are independent since  % se definisco non random tolgo tipically
loops that could induce such correlations, do not exist on  Bethe lattices,.  In other words, moving one of the neighboring particles away from a site
should not affect the ability of its other neighbors to move and an unfrozen neighboring
particle can itself move provided an appropriate set of particles that are further  up {\it its} branch of the tree have moved out of the way.
% If the number of particles in this set is finite, this will
%not affect the ability to move of the other neighbors (of the
%particle on which we are focusing on) because there are no finite
%loops. 
One can repeat the above argument to
show that the neighbors of $x$ must each have at least $m-1=z-s-2$ neighbors along their own
branches occupied by frozen particles,   or else all their unblocked 
neighbors must themselves have more than $z-s-1$ frozen neighbors,  and so on
and so forth.  
Therefore, the probability $\tilde P$ that a given site is occupied by a frozen particle can be expressed in terms of $Y$ and $B$
as

\begin{equation}
\tilde{P}=\rho \left(\sum_{j=k-s+1}^{k+1}\!\!Y^{j}(1-Y)^{k+1-j}\left(\genfrac{}{}{0pt}{1}{k+1 }{j}\right)
+\sum_{j=s+1}^{k+1}B^{j}Y^{k+1-j}
\left(\genfrac{}{}{0pt}{1}{k+1 }{j}\right)\right) \ .
\label{ptilde}
\end{equation}
As a consequence in the ergodic phase $Y=B=0$ and $\tilde{P}=0$,
whereas in the non ergodic phase $Y$ and $B$ are strictly positive and
thus so is $\tilde{P}$ \footnote{We expect that our derivation of
$\tilde{P}$ is correct under the hypothesis that to pick away a
particle during the bootstrap procedure, it is not necessary  to pick
away before so doing, an infinite (i.e. diverging with the system size) number of other
particles. This should be roughly equivalent to the hypothesis  that an appropriately defined correlation length is finite, a reasonable expectation   
 except at the bootstrap transition point. }.

This result leads to a physical interpretation of the dynamical
transition as a ``jamming" transition: at $\rho_{c}$, an
infinite set of blocked particles suddenly appears and concomitantly 
a breaking of ergodicity occurs. 

We have checked the relation between the  bootstrap percolation transition
and the dynamical transition by performing numerically the  bootstrap procedure for the case  $k=3, s=1$, and verifying that the density above which a cluster of  particles remains at the end of the process is compatible with the critical density for the dynamical transition.

\subsubsection{\bf Quantitative results for $k=3,\  s=1$ and $k=5, \  s=2$ models} \label{bethe23}
\label{k3}
We now consider in detail the  cases $k=3$, $s=1$ and $k=5$, $s=3$, which mimic, respectively, the non-trivial square lattice model, and the three-dimensional cubic-lattice model
originally introduced by Kob and Andersen.

In the case $k=2$ $s=1$, equations (\ref{p123}) give

\begin{eqnarray}
\label{pp123}
Y'&=&\rho \left(Y^3+3Y^2(1-Y)+B^3+3B^2Y\right)\nonumber\\
B'&=&(1-\rho)F^3
\nonumber\\
F'&=&\rho(Y+B)^3
\end{eqnarray}
from which one can immediately see that, at a fixed point with $Y>0$ also $B>0$ and  $F>0$. Therefore one can study the transition by analyzing the fixed point equations for 
\be
G\equiv Y+B
\ee
 which can be written in polynomial form:

\begin{equation}
\label{rhs}
 R(G)=G[-1+3\rho G-2\rho G^2+v\rho^3G^8-6v\rho^4(G^9-G^{10})+3v^2\rho^7(G^{17}-G^{18})]=0 \ .
\end{equation}
By direct analysis one can see that there is a fixed point with $G=0$ which is stable for {\it any} density, while above a critical density, $\rho_c$,  a second physical solution appears with positive $G$; both are stable in this regime. Separating these stable fixed points is an unstable one at an intermediate value of $G$.   As the density decreases to $\rho=\rho_c$, this unstable fixed point annihilates with the non-zero stable fixed point at  a saddle-node bifurcation signaled by 
\begin{equation}
\frac{\partial {R(G,\rho_c)}}{\partial G}=0
\end{equation}
at the fixed point.
\\
The values of the critical density $\rho_c$ and critical $G_c\equiv G(\rho_c)$ are:
%\begin{equation}
%\label{parcr}
%\rho_c \cong 0.888825 \qquad G_c\cong 0.755860
%\end{equation}
% a me vengono cosi'
\begin{equation}
\label{parcr}
\rho_c \simeq 0.888 \qquad G_c\simeq 0.758
\end{equation}
The critical density is strictly  less than, although only very slightly so, the
critical density $\rho_T=\frac{8}{9}$, of the much simpler
two-branching-subtree percolation problem discussed in section (\ref{existence}).

At $\rho=\rho_c$ the value of $G$ jumps discontinuously from the low
density value $G=0$ to $G_c$ and
then increases with a square root cusp: $G\approx G_c+C\sqrt{\rho-\rho_c}/\rho_c$, see figure \ref{boot1}.  
Similar behaviour obtains for the probabilities $\tilde P$ of  a site occupied by a frozen particle, as can  be obtained from equations (\ref{ptilde}) and (\ref{pp123}).

\begin{figure}[bt]
\psfrag{G}[][]{{\large{G}}}
\psfrag{rho}[][]{{\large{$\rho$}}}
\psfrag{logG}[][]{{\large{$log (G-G_c)$}}}
\psfrag{logrho}[][]{{\large{$log(\rho-\rho_c)$}}}
\centerline{
\includegraphics[angle=-90,width=\columnwidth]
       {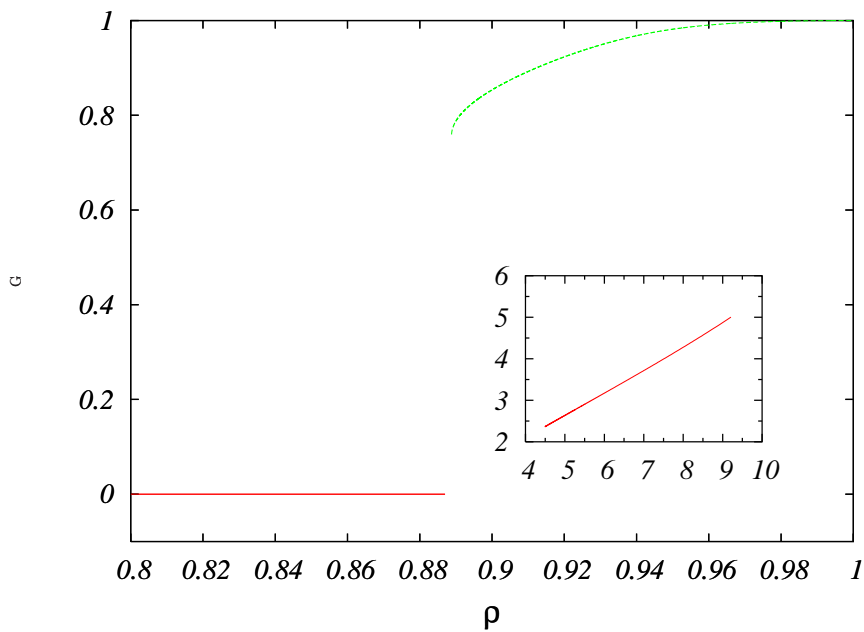}}
\caption{The probability, $G$ (defined in the text) as a function of density for  $k=3$, $s=1$ Bethe lattice as obtained  from the stable
  solution of equations (\ref{rhs}). In the inset 
  $log(G-G_c)$ is shown as a function of $log(\rho-\rho_c)$ in the vicinity of
  $\rho_c$ showing the square root dependence.
\label{boot1}}
\end{figure}

 For the case $k=5$, $s=2$, by solving equations (\ref{p123}),
a similar  transition is found at a slightly higher critical density 
$\rho_c\simeq 0.915$ 
%(see figure \ref{boot3}).
\subsubsection{Diverging time and length scales at the dynamical transition}\label{scales}
Although the dynamical transition in these KA models is discontinuous,
there is precursor behavior more characteristic of critical
transitions.  In this respect, the situation is somewhat analogous to
conventional bond percolation on one dimensional lattices with
connection probability between any two points proportional to the
inverse square of the distance between them \cite{lr-perc}.

Indeed from the discussion in the previous subsection, an interpretation of the mixed
nature of  the dynamical  transition of Bethe lattice KA models is suggested:
the dense cluster of frozen particles that arises immediately at $\rho=\rho_c$
contains a finite fraction of the particles and it is, at the same time,  {\it fragile} in a way similar to incipient infinite clusters at conventional critical percolation.   This fragility implies that by removing a  tiny, but carefully chosen,  fraction of the particles, the frozen  cluster can be ``melted" and will disappear.
Associated with this fragility is a correlation length  that diverges
as  $\rho\to\rho_c$ from above. The form of the divergence,  
$\sim (\rho-\rho_c)^{-\frac 12}$ is controlled by the ``rate" - in
tree levels -  of approach of the iterated probabilities to the fixed
point. 
Physically, this correlation length is the (chemical) distance
beyond which it is unlikely to have to move vacancies  in order for a given  
particle to be moved.

More significant than the long length scales that appear near the critical density, are the corresponding  dynamical effects.  In particular, we expect a diverging time scale as $\rho_c$ is approached from below. Unfortunately, our analytic methods, while they take into account the dynamical constraints, cannot directly be used to compute the dynamical correlations of primary interest.  At this point, we resort to numerical simulations to study these and other properties of the Bethe lattice models. 

Before turning to the numerics, we note that the behavior on  Cayley trees, because of their boundaries which account for a positive fraction of the volume,  might exhibit  different  phase diagrams depending on the choice of boundary conditions.

% [If the transition at which disturbances can be propagated to infinity also occurs at $r_c$, there may be another length scale associated with this. If the transition occurs at a higher value of $r_c$, there will already be an infinite length for a range above $r_c$. If the transition has not occured yet at $r_c$, there may only be the one divergent length scale.]
%Below but near the transition, there will be a large characteristic
%length scale which marks the crossover from the tree tending to have
%many frozen particles to tending not to.  Associated with this, there
%should be a diverging characteristic length scale and, presumably, a
%diffusion coefficient that vanishes as the transition is approached
%from smaller $\rho$.\\

\subsection{Dynamics and numerical simulations}
\label{bethedynamics}

In order to understand  in more detail the character of the dynamical
transition, we have performed numerical simulations of KA dynamics
for  the $k=3$ $s=1$, model on a random c-regular graph  of $N=10^4$
sites with coordination number $k+1=4$.

In particular we have computed
the  local density-density dynamic
auto-correlation function $C(t)$ and the corresponding dynamical
susceptibility $\Psi  (t)$:
\begin{eqnarray}
C(t)&=&\frac{1}{N}\sum_{i=1}^{N}\frac{\langle n_{i}(t)n_{i}(0)\rangle -\rho ^{2}}{\rho -\rho ^{2}}\nonumber\\ 
\chi_4(t)&=&N\big\langle\left(\frac{1}{N}\sum_{i=1}^{N}\frac{n_{i}(t)n_{i}(0)-\rho
^{2}}{\rho -\rho^{2}}-C(t) \right)^{2}\big\rangle\nonumber
\end{eqnarray}
where $\langle \ \rangle$ denotes an average over the  Bernoulli product measure at
density $\rho$ for the choice of the initial configuration at over the Montecarlo dynamics. 
The dynamical susceptibility 
$\chi_4  (t)$ (and its generalization) 
has been introduced in \cite{FranzDH} and is currently
used \cite{Glotzer,Berthier} to detect a growing 
dynamical correlation length in simulations of glass-forming liquids.

\begin{figure}[bt]
\centerline{    \epsfysize=8cm
       \epsffile{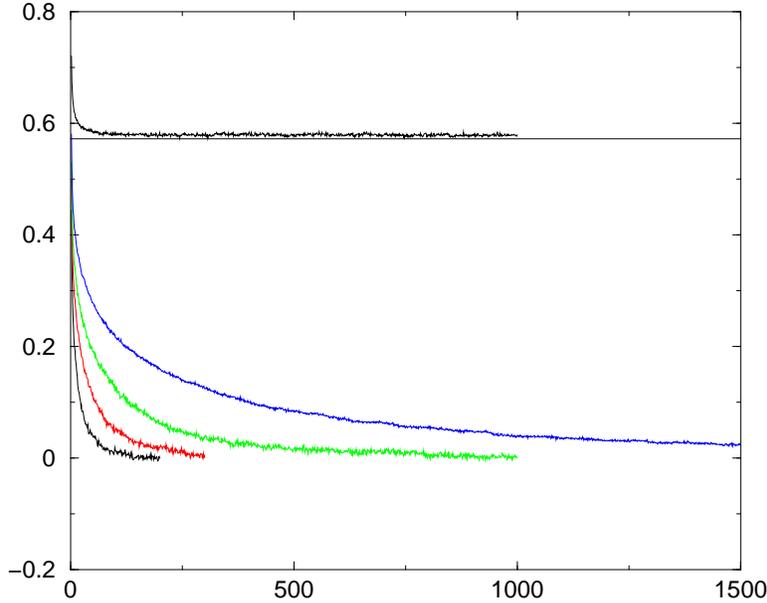}}
\caption{Dynamic autocorrelation function $C(t)$ as a function of time 
 for densities $0.85,0.86,0.87,0.875,0.9$ (from down to up) on a 10000-site approximation to a Bethe lattice with coordination number $k+1=4$ and single-vacancy assisted dynamics.
The dynamical transition is clearly evident between the densities $0.875$ and $0.9$. The straight
line is the value of the Edwards-Anderson order-parameter obtained by the approximation discussed in the text.
\label{figC}}
\end{figure}

 \begin{figure}[bt]
\centerline{    \epsfysize=8cm
       \epsffile{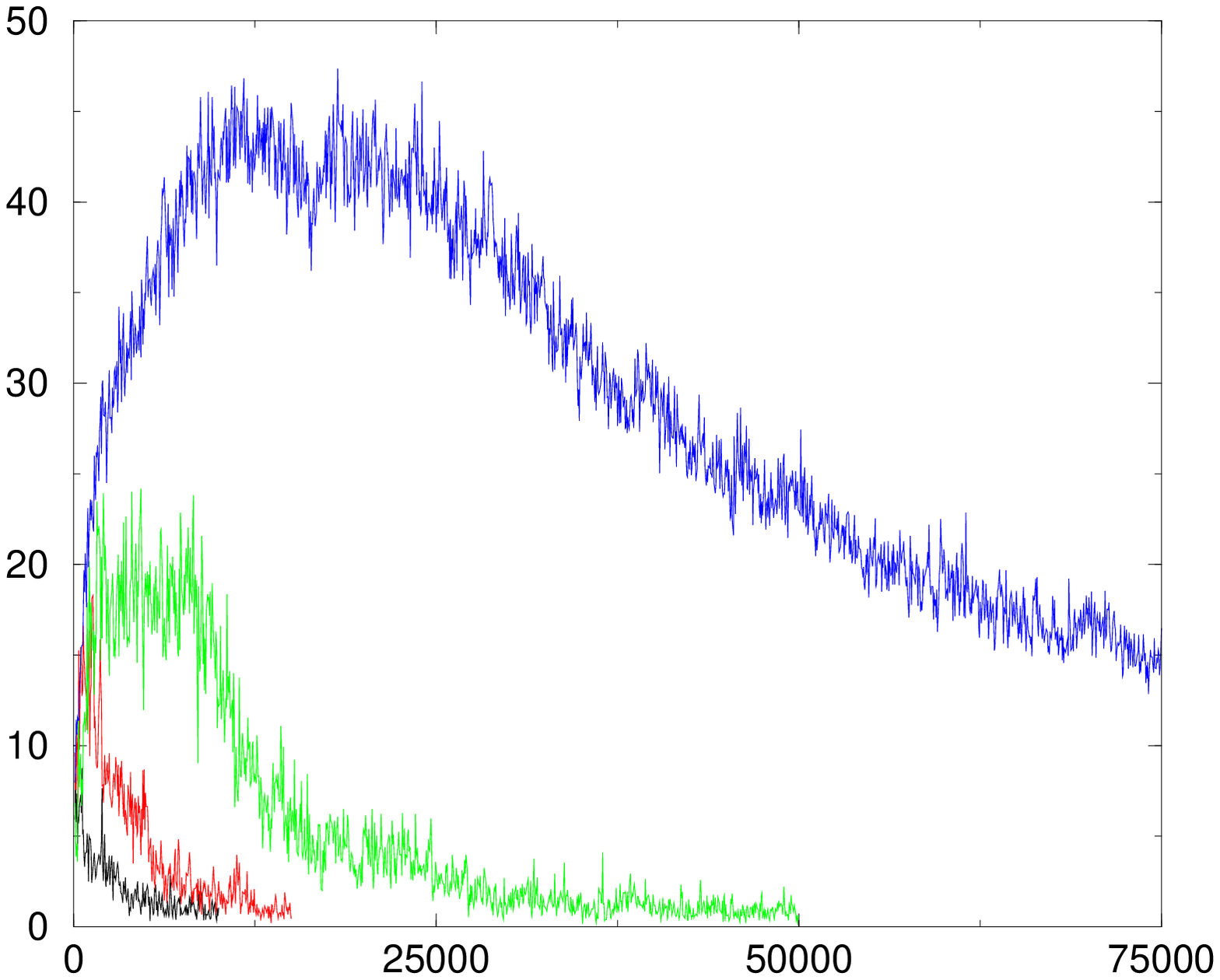}}
\caption{Dynamical susceptibility $\chi_4 (t)$  as a function of time 
for densities $0.85,0.86,0.87,0.875$ (from down to up) on a 10000-site approximation to a Bethe lattice with coordination number $k+1=4$ and single-vacancy assisted dynamics.
\label{figPsi}}
\end{figure}

The results for $C(t)$, plotted in figure \ref{figC}, show that the equilibration time grows and seems to diverge at the critical density found from the quasistatic analytical calculations (see equation (\ref{parcr})). Indeed, for $\rho <\rho_c-\Delta$, $C(t)$ tends  to zero at long times but with a characteristic decay time that  increases at  higher densities. Here $\Delta$, the precision obtained for $\rho_c$, is $0.0075$. For $\rho <\rho_c+\Delta$, $C(t)$ no longer appears to decay to zero but seems to approach a non-zero plateau whose magnitude it is natural to  call an Edwards-Anderson  parameter by analogy with spin glasses. 
%the p-spin case. 
.
Since at the critical density there should be $N\tilde P(\rho_c)$
frozen  particles, the value of the Edwards Anderson parameter
$q_{EA}\equiv\lim_{t\to \infty} \lim_{N\to \infty} C(t)$ has two
contributions coming from sites that are occupied respectively by a frozen particle
or by a vacancy that can never move. These can be
computed exactly from the quasi-static computation developed in the
previous section. However there is a third contribution that
corresponds to 
\be
\lim_{t\to\infty}\frac{1}{N}\sum_{i=1}^{N'}\frac{\langle
n_{i}(t)n_{i}(0)\rangle-\rho ^{2}}{\rho -\rho ^{2}} \nonumber
\ee
 where the index
$i=1,\dotsc ,N'$ runs only over the sites whose occupation number changes
during the dynamics (i.e. they are not occupied by vacancies or
frozen particles). This contribution to $q_{EA}$ is more cumbersome to compute. 
In this paper we  just give a rough approximation which consists
of assuming that the long-time limit of $\langle
n_{i}(t)n_{i}(0)\rangle$ on these sites --- which consist of many small disconnected clusters --- is equal to the density of
particles within the full set of $N'$ dynamic sites. This can be computed from 
the quasi-static analysis  developed in the previous section.
Comparison with numerics shows that this approximation works rather
well (see e.g. Fig. \ref{figC}). 

A dynamical  quantity that can be computed exactly is the long-time limit of  the {\it  persistence function}: the probability that the
occupation variable of a given sites has not changed from its initial 
value for an interval of time $t$. The long-time plateau of this, by definition,
is equal to the fraction of sites that are occupied by either a frozen particle
or by a vacancy that can never move.

Plots of the dynamical susceptibility, $\chi_4(t)$, found in the
simulations are shown in figure \ref{figPsi} 
for various densities: these appear consistent with a
divergence at the critical density. More precisely 
both  the height of the peak and the characteristic time scales at which this occurs, appear to diverge at $\rho_c$.

%Note that in principle there could have been a lower critical density, below the $\rho_c$ at which an infinite cluster of frozen particles appear,  at which  the dynamical susceptibility or the relaxation times, diverged.  The results from the simulations indicate that this  is not the case: as for conventional percolation,  as the density increases the divergence of the associated susceptibility is immediately followed by the appearance of an infinite cluster. 

%\begin{figure}[bt]
%\centerline{
%\includegraphics[width=\columnwidth]
% {ptildecor.eps}}
%\caption{$\tilde{P}$ as a function of density for $k=3$, $s=1$. Dots are from the solution of equations (\ref{ptilde}) and (\ref{pp123}); solid line is the curve $P_{c}+(\rho-\rho_c)^(1/2)$ with $P_c\simeq 0.596$ which gives the right behavior in the vicinity of $\rho_{c}$ from above.}
%\label{boot2}
%\end{figure}
%
%\begin{figure}[bt]
%\centerline{
%\includegraphics[width=\columnwidth]
%       {ptilde5.eps}}
%\caption{$\tilde{P}$ as a function of density for $k=5$, $s=2$. Dots are from the solution of equations \ref{ptilde} and \ref{p123}; solid line is the curve $P_{c}+(\rho-\rho_c)^{1/2}$, $P_c\simeq 0.725$ which gives the right behavior in the vicinity of $\rho_{c}$ from above.}
%\label{boot3}
%\end{figure}

\begin{figure}[bt]
\centerline{ 
\includegraphics[width=\columnwidth]
       {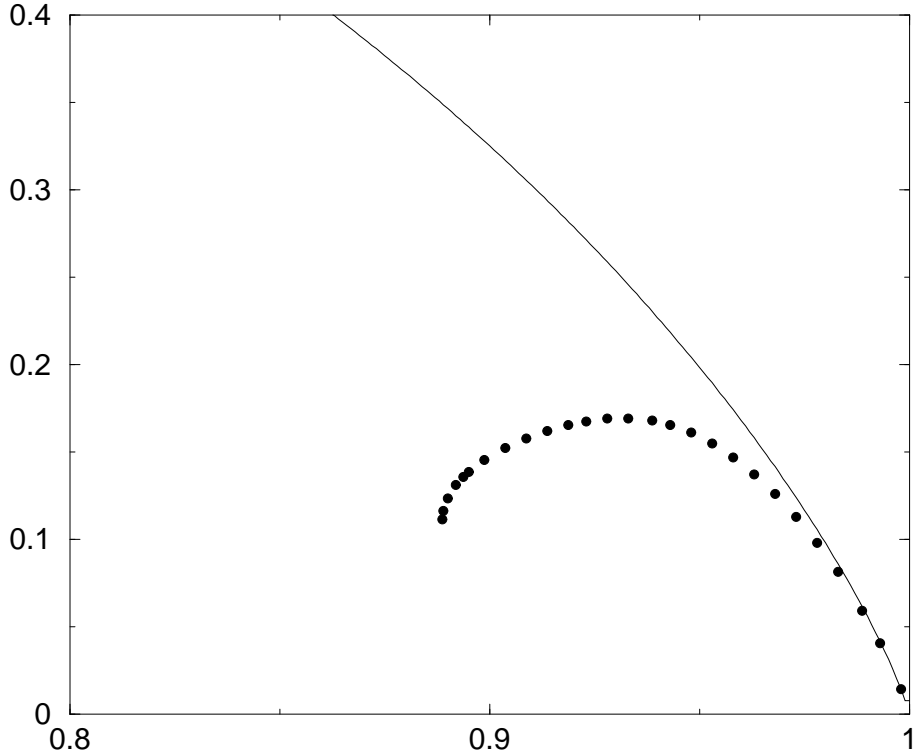}}
\caption{ Lower bound, $S^b(\rho)$, on the configurational entropy --- logarithm of the number of frozen configurations per site --- as a function of density for the $k=3$  $s=1$ Bethe lattice is indicated by the dotted line; this bound is obtained from equation (\ref{bound}) and numerical solution of equations (\ref{ptilde}), (\ref{pp123}). Below the critical density $\rho_c\simeq 0.888$, the lower bound jumps to zero. The straight line is the equilibrium entropy.
\label{figS}}
\end{figure}

\subsubsection{\bf Configurational entropy} 
\label{configurationalentropy}

If phase space breaks into a sufficiently large number of different ergodic components, some of the equilibrium {\it entropy density} will be associated with the logarithm of the number, ${\cal{N}}_e$, of distinct components:
\be
S_c(\rho )=\ln{\cal{N}}_e/V
\ee
where $V$ is the total number of sites.  This is referred to in the glass
literature as the {\it configurational entropy} \footnote{This terminology is somewhat 
misleading since in general  statistical mechanics  
``configurational  entropy" refers to the sum over all configurations (after having
integrated out the momenta). For glasses, its meaning is the
contribution to the entropy density coming from the large number of
different ergodic components. The other contribution to the entropy is
related to the number of configurations within the given ergodic component.}.

For KA models on Bethe lattices  one can find a lower bound for the configurational entropy 
by observing that
 all the configurations belonging 
to the same ergodic component must have the same cluster of frozen particles, therefore ${\cal{N}}_e$ 
is at least as large as  the number
of different sets of frozen particles, ${\cal{N}}_f$, which has  an entropy density,
\be
S^{i}(\rho ,\rho _{F}) = \frac{1}{V}\ln{\cal N}_f \ . 
\ee
Furthermore, the entropy of the non-frozen sites, $S^{i}(\rho ,\rho_F)$, can be bounded above by the number of ways of putting $V\rho-V\rho_F$ particles on $V-V\rho_B$ sites, where $\rho_F$ is the density of frozen particles at total density $\rho$:
\begin{equation}
S^{i}(\rho ,\rho _{B})\leq (1-\rho _{B})\ln (1-\rho _{B})-(\rho -\rho _{B})\ln(\rho -\rho _{B})
-(1-\rho )\ln (1-\rho )  \  .
\end{equation}

Since the total entropy density,
\be\label{entropy}
S(\rho)=-\rho\ln\rho -(1-\rho)\ln(1-\rho)=
\left( S^{f}(\rho ,\tilde{\rho _{F}})+S_c(\rho )\right) \ ,
\end{equation}
and we know that $\tilde{\rho_F}=\tilde{P}$ that was computed earlier,
see eq. (\ref{ptilde}), we obtain a lower bound for the  configurational entropy
\begin{equation}
\label{bound}
S^c(\rho)\geq S^f(\rho,\tilde P)\geq -\rho \ln \rho -(1-\tilde P)\ln (1-\tilde P)
+(\rho -\tilde P)\ln (\rho -\tilde P) \ .
\end{equation}
In figure \ref{figS} this lower bound is plotted as a function of density for the case $k=3$, $s=1$.
Note that, among other factors, we have ignored in this estimate the existence of empty but blocked sites: the effect of these will be to lower the entropy of the unfrozen particles.

%Furthermore, by means of a saddle point evaluation,
 %the entropy at the density $\rho $ can be expressed as

%\begin{equation}

%S(\rho )=forever blocked particles for an equilibrium confuguration at density $\rho$, $S^{i}(\rho ,\rho _{B})$ is $1/N$ times the logarithm of the number of configurations with the same cluster of $N\rho _{B}$ particles blocked forever and $S^{c}(\rho ,\rho _{B})$ is $1/N$ times the logarithm of  ${\cal{N}}_b$.
%An upper bound on $S^{i}(\rho ,\rho _{B})$ can be readily calculated by considering that $\exp{NS^i(\rho,\rho_B)}$ is smaller than  therefore

%where we have used Stirling formula.
 %Since
 %$S(\rho )=-\rho \ln \rho -(1-\rho )\ln (1-\rho )$ and
 %the value of $\rho _{B}$ 
%which
%maximizes (\ref{entropy}) is the typical one, i.e. 
 %$\tilde{P}$,

In conclusion, our results from lower bounds and analytical and numerical
analysis of the dynamical transition  support the expectation that the configurational entropy
jumps form zero  to a non-zero finite value precisely at $\rho_c$. For $\rho>\rho_c$, the configurational entropy decreases with the decreasing vacancy density and overall entropy; it 
vanishes at $\rho =1$. 
%The maximum configurational entropy is thus attained right at the transition. 

\subsection{\bf Bethe lattices with loops}
\label{betheloops}

%vedi Monroe j.s.p 65 91 p225. 67 '92 p.1185 per husimi tree e essam fisher rev mod. phys 42 p.272 '70

In order to investigate whether the absence of loops is the essential reason for the existence of a phase transition on Bethe lattices, we briefly consider structures that are tree-like on large scales, but do have loops. 

Specifically, consider the rooted tree composed of triangles with one  vertex pointing downwards and the other two vertices each being the bottom vertex of another triangle and  hence the root of a tree.   Then define a  Bethe lattice with triangular loops as the graph obtained by taking two such lattices and merging the free vertices of the two roots (see figure \ref{figloops}). The coordination number of such a tree is $z=4$ and the branching ratio $k=2$.
This is an example of a cactus, or Husimi, tree  \cite{cactus} with primary clusters of three vertices and  three branches departing from each cluster.   
 \begin{figure}[bt]
\centerline{ 
\includegraphics[width=0.25\columnwidth]  
       {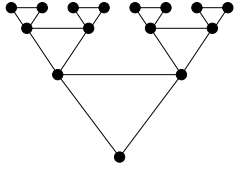}}
\caption{A branch of a Bethe lattice with triangular loops. 
\label{figloops}}
\end{figure}

Consider the KA model with $s=1$ on the triangular tree defined
above. It is useful to focus on the set  ${\cal{E}}_i$ of
configurations of the subtree rooted at  triangle $i$, such that the bottom vertex is occupied and a 
vacancy  cannot be brought down to it  without using
vacancies below it on the tree. 
Let $P$ be the probability of such an event. If a configuration, $\eta$
has all three sites of triangle $i$ occupied and all  of the  subtrees rooted on  triangles two
levels up from $i$,  $i^1,\dots,i^{4}$, have configurations $\in
{\cal{E}}_i^j$, then the configuration of the subtree rooted at triangle $i$  belongs to ${\cal{E}}_i$.
The same is true for any configuration, $\eta$,
that has the three sites of $i$ all occupied, and $\eta\in {\cal{E}}_{\tilde i^j}$  for three of the four subtrees rooted at the $\{i^j\}$.  Therefore 

\begin{equation}\label{Pineq}
P\geq f(P)=\rho^3\left(P^4+4P^3(1-P)\right) \ .
\end{equation}
 Since $P=1$ is a fixed point of $P'=f(P)$ when $\rho=1$ ,
 at sufficiently large density there exists 
a $\bar{P}>0$ such that $f(\bar{P})=\bar{P}$.
If one iterates down the tree for  actual $P$s, then the
inequality (\ref{Pineq}) implies that, starting from an initial
condition $P_0=\bar{P}+\epsilon>\bar{P}$, all the
subsequent $P$'s, $P_1,P_2,\dots$, will be larger than $
\bar{P}$. Indeed, since the $f(x)$ is monotonically increasing
in $0<x< 1$, $P_1\geq f(\bar{P}+\epsilon)>f(\bar{P})=\bar{P}$ and  the
same is true at any subsequent step.  Therefore, the full fixed
point must have $P^*>\bar{P}>0$. 
Thus for sufficiently large $\rho$, $P^*(\rho)$ will be non zero and the
system in a frozen phase. 
It immediately follows that the density of frozen particles, $\tilde{P}$ is non-zero.

A generalization of the argument in section \ref{existence} allows one to prove that there exists a transition at a non-zero density from a regime where $P=0$ to a region where $P$ is positive. %More naturally, the critical density can be defined as the value of density at which  the probability 
%$\tilde P$ that a given site is occupied by a particle blocked forever goes from 
%zero to a finite value. 
Let $\overline{\cal{E}_i}$ be the complement of ${\cal{E}}_i$ defined
above.  A configuration is definitely in $\overline{{\cal{E}}_i}$  if
the four branches rooted at the four triangles two levels up from $i$
are in $\overline{{\cal{E}}_{i^j}}$. The same is true if
 just three of them are in $\overline{{\cal{E}}_{i^j}}$ and furthermore
one can have or
 bring vacancies in the highest sites of the triangle which is just above the
site which is in the same triangle of the site that cannot have the vacancy.  
This sufficient condition for $ \overline{\cal{E}_i}$ yields and upper bound on $P$:

\begin{equation}
P\leq 1-(1-P)^4-4P(1-P)^4
\end{equation}
which implies --- as in section \ref{discontinuity} --- that $P$ has a {\it jump} from zero to a non-zero value  at the critical density.

The arguments given above can be extended to KA models with any constraint parameter, $s$, on generalized Bethe lattices with $n$-polygonal loops for any $n$, i.e. trees composed of polygons of $n$ vertices, each which is also a vertex of another polygon.
To prove the existence of the transition, it is sufficient to consider
the set of configurations  ${\cal{E}}_i$ such that a
vacancy is not and cannot be brought down to the bottom vertex $i$ without using
vacancies below. By the same argument as above, we can bound the probability $P$ of such an event by
\begin{equation}
P\geq\rho^n\left(P^{(n-1)^{2}}+ (n-1)^{2} P^{(n-1)^{2}-1}(1-P)\right)
\end{equation}
which implies again that $P>0$ for sufficiently high density.
%To prove that the transition is discontinuous for $s<k-1$, we focus on the less restrictive event that the bottom site of a polygon is occupied and the particle cannot move as long as the other $n-1$ sites of the polygon below it are occupied as well. The probability $R$ of this event, for $s<k-1$ can be bounded as
Generalizing the procedure used for the triangular tree one obtains an bound in the opposite direction:
\begin{equation}
P\leq 1-(1-P)^{(n-1)^{2}}- (n-1)^{2} P^{(n-1)^{2} -1}(1-P)
\end{equation}
which implies, as above, that the transition is discontinuous.

Another class of Bethe-like lattices with loops is also instructive to
analyze. 
Consider, as an  example, starting with  normal Bethe lattice with connectivity four and
replacing each site by a square lattice $\Lambda\in\bZ^2$ of linear size L connected to to other squares along its edges (see
figure \ref{newbethe}).
If we  arrange the lattice as a tree with one branch pointing  down from each vertex, the ``upwards" branches from a square consist of an upwards extension of the lower branch and side branches emerging perpendicularly from the other two sides. Locally, this looks like a square lattice except at  the corners: the corner sites still have four neighbors --- albeit on three other squares --- but have eight second neighbors that are each on separate ``sheets" in contrast to the four second neighbors on a regular square lattice each of which can be reached in two distinct ways.

\begin{figure}[bt]
\psfrag{a}[][]{(a)}
\psfrag{b}[][]{(b)}
\psfrag{c}[][]{(c)}
\centerline{
\includegraphics[width=\columnwidth]{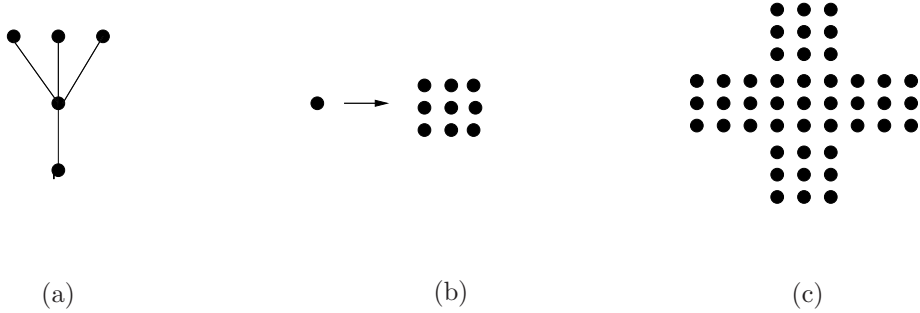}}
\caption{ Bethe lattices with loops (c) obtained by substituting sites with
  squares of size $L$ (b) from a 
  normal Bethe lattices with connectivity four (a). Here $L=3$. 
}
\label{newbethe}
\end{figure}
Drawing the tree of squares as described above immediately implies that for the $s=1$ KA model on which we will focus, there exist infinite networks of frozen particles.As for the normal square lattice, these consist of two-wide fully occupied bars, each either extending to a faraway boundary, or beginning and terminating at T-junctions with other such bars.  But such a network, because of the large scale topology of the Bethe lattice, is very different from its real-square-lattice counterpart.  It is still very fragile but there are so many potential T-junctions, which are independent of each other if the bars emerging from them go onto different branches, that such a frozen network exists almost surely at sufficiently high densities. The reason for this is  simple: consider a vertical bar coming into a square, $i$, from below. There are at least two possibilities: the bar can continue through $i$ and into the square above it, or it can end at a T-junction with a bar that extends horizontally into {\it both} of the other two neighboring squares of $i$.   Since the probabilities, $T$, that the extensions into each of the three neighboring squares exist and are themselves part of infinite frozen networks on their own sub-trees --- i.e., excluding square $i$ ---  are independent of each other, one can bound from below the probability that square $i$ has a bar rising vertically from its lower boundary and either extending through the top boundary or ending at a T-junction with a bar that extends into the two side boundaries, in terms of $T$ and $\rho$.  As in cases we have already analyzed, this can be used to show that there must be an infinite frozen network at sufficiently high densities. One can likewise  show that the transition to the frozen phase will be discontinuous.

To show that the transition on this  Bethe lattice of squares is discontinuous, it is useful to focus on the framing by vacancies of individual squares, in the sense used for the conventional square lattice.
% Then
%consider a square,  $I$,
%and focuse on the event that there are no vacancies at the verteces of $I$ and, without taking advantage of the
%configuration in the bottom square, one cannot bring any vacancy on such vertices. Let $P$ be the probability of such event. If
% all the $L^2$ sites of square $I$ are filled and at least two of the
%squares above have all the vertices always occupied, than  $I$
%will have
%for sure all the vertices occupied.
%Therefore 
%\begin{equation}
%P\geq \rho^{L^2}\left(P^3+3P^2(1-P)\right)
%\end{equation}
%and, by an argument analogous to the one in section \ref{existence}, the existence of a transition
%follows. 
Consider  the event that, without taking advantage of possible vacancies in the square below $I$, one cannot 
frame the configuration in $I$,
i.e. reach a configuration with all the boundary sites of $I$ empty.
Call $X$ the probability of such an event.
If, without using the square $i$,  the three squares left, right and above $I$  can all framed, or if either 
the above and left or the above and right squares can be framed, or if
the left and right squares can be framed and in the 
square $I$ the first row is empty and all the others are occupied
then the square $I$ can definitely be framed. These possibilities are sufficient, but not necessary. Therefore

\begin{equation}
X \leq 1-(1-X)^3-[2+\rho^{L^2-L}(1-\rho)^L] X (1-X)^2
\end{equation}
which is incompatible with a continuous transition by an argument similar to that used for other Bethe lattices.

In general, the existence of finite loops on these decorated Bethe lattices, makes the motion of particles less restricted that on a regular Bethe lattice. Indeed, there will be configurations in which  two or more vacancies can be ejected from one branch up into another, and by making use of these, trigger 
the potential ejection of a larger number of vacancies down from the second branch. This process can depend on configurations arbitrarily far up in the two branches.  Such  ejected vacancies could in principle extend their influence arbitrarily further down the tree and render all  particles mobile, even when the density is above the critical value for the loopless Bethe lattice.   But the above results show that for low enough vacancy concentration this will not occur: the system will be in a partially frozen state like that of the simple Bethe lattices. The above picture does, however, suggest a route to estimating the critical density as a function of $L$.

The trees-of-squares we have been studying  interpolate between simple Bethe lattices, for $L=1$, and a full square lattice for $L=\infty$. Therefore on the trees-of-squares, in the limit $L\to\infty$ the critical density, $\rho_c(L)$ should go to one.
But {\it how} does it approach one?  From scaling arguments, one would
guess that the framing probabilities on the tree will be functions of the ratio of 
$L$ to one of the characteristic lengths of the infinite
square system: i.e. functions of  either $L/\xi(\rho)$, or $L/\Xi(\rho)$.
Naively, one might expect that the behavior would be controlled by the large crossover length, which
determines the spatial density, $1/\Xi(\rho)^2$, of mobile cores and thus the
probability that squares of size $L$  are frameable. However, this ignores the effects of possible help
from vacancies moved from other squares and the correct behavior can be seen
as follows. Consider one of the $L$ by $L$ squares that {\it can} be framed by
vacancies.  The probability that this framing can be continued all the
way through a given one of its neighboring squares to make an $L\times
2L$ framed rectangle, is of order $[a (1-\rho )L]^L$ with an order-one coefficient $a$.  This will be substantial as long as $\xi(\rho)>L$ --- in spite of  the fact that the probability the original square was frameable is very small unless $L\sim \Xi(\rho)$.  The crucial observation that leads to a scaling with $\xi$, is a consequence of the tree structure: once the neighbor square of $i$ has been framed together with $i$, as above, each of the three other neighbors of this neighbor can potentially join the frame as there already exist complete rows of vacancies along one of their sides from which the framing can be extended.  Since any of these squares that joins the frame gives rises to another three potential squares that might join as well, if the original square frame can be extended out in several directions to make a small frameable tree, it become likely that it can be extended much further, indeed, to infinity.  Thus the crucial condition for extending the framing to infinity is that there be a substantial probability of a frame around one square being extendable across {\it one} of its neighboring squares --- even when the probability that a single square is  frameable on its own is beery small. 

We conclude that the critical density, $\rho_c(L)$ for the $s=1$ KA model on a tree of $L\times L$ squares, will be given by 
\be
\xi(\rho_c)=bL
\ee
with some order-one coefficient $b$.  As $\xi\sim\ln[1/(1-\rho)]/(1-\rho)$ for high densities, this yields
\be
\rho_c(L) \approx 1-\frac{B\ln L}{L}
\ee
with some coefficient $B$.  Note that this form is the same as that suggested by the argument given above for the appearance of an infinite network of bars of frozen particles, since the probability of a two-wide fully occupied bar crossing a square   is of order $L\rho^{2L}$ which is of order one at $\rho_c(L)$.

\section{\bf Fredrickson Andersen model}
\label{FA}

The Fredrickson Andersen (FA) model is a spin model with  kinetically
constrained dynamics that is similar to the KA model but without a conservation law: roughly speaking
KA models are FA models with dynamics that conserves the number of
particles. Some of the results discussed previously for  KA models have been
previously obtained for  FA models \cite{FA,Reiter,ReviewKLG}: for example
irreducibility on some hypercubic lattices for the non-trivial range of the constraint parameter . However, there are
important results, like the absence of a ergodic/non-ergodic
transition, that have not been proven previously, as well others,
such as  the dependence of the relaxation time scale on density, that
can be made sharper. In the following we briefly discuss how these
new results on  FA models can be obtained using the same techniques 
that we have developed for KA models. 

\subsection{\bf Definition of the model}
\label{FAdef}

Let $\L$ be an hypercubic $d$--dimensional lattice and $f$ an integer parameter in the range $0,..2d$.
The FA model \cite{FA} is a facilitated Ising model with ``occupation" variables $\eta_x\in\{0,1\}$ --- configuration space  $\O_\L=\{0,1\}^{|\L|}$.  {The FA model is conventionally defined in terms of spin variables; we use 
occupation variables to make comparisons with KA models more natural.]  The occupation variables can ``flip" with rate

\begin{equation}
\label{ratesFA}
 c_{x} (\eta) :=
\left\{
\begin{array}{ll}
0 & \textrm{ if} \quad n_x(\eta)\leq f \\
% & \textrm{  \ \ \ \  that are empty is  less than
%\ } \displaystyle{
%\sum_{\genfrac{}{}{0pt}{1}{d(\{z\},\{x\}) =1}{z\in\L}}
%f}\\
%(1-\eta_z)<f\\
{\mbox {min}}\left (1,\exp{-\frac{\Delta H}{T}}\right) & \textrm{ otherwise \ }
\end{array}
\right.
\end{equation}
where 
\be
n_x(\eta) := \sum_{ \genfrac{}{}{0pt}{1}{y\in\Lambda}{|x-y|=1}}(1-\eta_y)  \ \ \ \ ,
\ee
i.e.  $n_x$ is the number of empty neighbors of $x$ and
the change in energy  $\Delta H=H(\eta^x)-H(\eta)$ comes from the non 
interacting Hamiltonian
\begin{equation}
H=-\frac h 2 \sum_x\eta_x
\end{equation}
with $h$ a positive coefficient that determines the fraction of occupied sites in equilibrium.
These rates satisfy detailed balance with respect to the trivial equilibrium 
product measure 
\begin{equation}
\mu_{\L,T}(\eta)=\prod_{x\in\L}
\left(\frac{\exp (\beta h)}{1+\exp (\beta h)}\right)^{\eta_x}
\left(\frac{1}{1+\exp (\beta h)}\right)^{1-\eta_x}
\label{gibbs}
\end{equation}
with $\beta=1/T$: this corresponds to Bernoulli measure with density  
\begin{equation}\label{gibs}
\rho = \exp (\beta h)/(1+\exp (\beta h)) \ .
\end{equation}
In contrast to KA models,  $\sum_{x\in\L}\eta_x$ is not conserved in FA models:
a  ``flip" at a single site
 corresponds to the birth or death of a particle.
 
In the unrestricted case $f=0$, the FA dynamics is simply Glauber dynamics for uncoupled Ising spins in a magnetic field: occupied sites corresponding to up spins and vacant sites to down spins. But for positive $f$,  a site can change  only if at least $f$ of its nearest neighbors
 are empty (i.e. their occupation variables equal zero), we refer to empty sites as
{\sl facilitating} and to $f$ as the {\sl facilitation parameter}. For $f>d$, fully occupied hypercubes will be frozen.  We thus restrict consideration to $1\le f\le d$.
Since energy favors occupied sites, 
in the low temperature regime most potential flips will not be allowed. Therefore we expect dynamics to be slow in
this regime, as
for the high density regime of KA models. 
On finite lattices, as for KA models,  there  is not a unique invariant measure: 
there exist  blocked configurations  and the FA dynamics are  not
ergodic. But irreducibility in the thermodynamic limit for $f\le d$ has been proved
for two and three--dimensional hypercubic lattice FA models,
\cite{Schonmann,Reiter,AizenmanLebowitz} (see \cite{ReviewKLG} for a review).

There is considerable similarity between KA and FA models: in the former, a
particle cannot move if it has more than $m$ occupied neighbors; in the latter
it cannot flip to vacant if it has more than $2d-f$ occupied neighbors. Thus
$f$ plays a similar role to $2d-m$, or, equivalently, to $s+1$, with $s$ the
KA model parameter that restricts a vacancy from moving if it has fewer than
$s$ neighboring vacancies.  In addition to their conservation law, the KA
models differ from FA models because of their restrictions on the neighbors of
two sites (the initial and final sites for the jumping particle) for each possible move.

\subsection{\bf Irreducibility and ergodicity}
\label{FAergodicity}

The presence or absence of frozen configurations in FA models is simply related to bootstrap percolation. Consider a configuration of occupation variables and perform a boostrap procedure by
removing  at each step all particles that have 
at least  $f$ empty neighbors. 
If, at the end of this procedure no particles  remain,
 the initial configuration belongs to the
irreducible component that  contains the totally empty configuration
(called the high temperature partition by
Fredrickson and Andersen).
This is exactly conventional bootstrap percolation. But all the ``moves" made in this bootstrap procedure are allowed in the corresponding FA model.
Therefore, using the results for the absence of an unpercolated phase  for bootstrap percolation \cite{Schonmann,AizenmanLebowitz}, one can
immediately conclude that in the thermodynamic limit irreducibility holds for FA models at any temperature as long as $f\le d$. In contrast to KA models for which bootstrap moves would not be allowed, for FA models bootstrap results are sufficient
to conclude irreducibility in the thermodynamic limit.  The result proven for the two and three--dimensional cases \cite{Reiter} is trivial to generalize to higher dimensional cases
using the  bootstrap percolation results \cite{cerfcirillo}.
Furthermore, the crossover length $L^{B}(\rho)$ that characterizes the bootstrap procedure is not only a lower bound for the crossover length, $\Xi(T)$ of FA models,  (as was the case for  KA models)  but
coincides with it under  the change of variables $\rho\to\exp (\beta h)/(1+\exp (\beta h))$. 

As we explained previously, irreducibility in the thermodynamic
limit is not a sufficient condition for ergodicity, which is the physically
interesting property for dynamics. However,
 ergodicity can be proven as in section
\ref{irreducibility}, using again irreducibility and the product form of equilibrium measure.

\subsubsection{Transition on Bethe lattices}

On Bethe lattices, FA models with positive $f$ displays a dynamical transition and aging dynamics
\cite{AndersenMCT,SellittoBiroliToninelli}.
This is a natural expectation that follows from the  bootstrap percolation transition on 
Bethe lattices:   our analysis for the KA model --- part of which involved removing particles as allowed in FA models --- 
leads, in the FA case on a Bethe lattice, to exactly the same bootstrap percolation equations as in \cite{Leath}. 
We will not discuss further Bethe lattice FA models.

\subsection{\bf Dynamics}
\label{FAtime}

We now turn to the dynamics for 
FA models on  hypercubic d--dimensional
lattices. 

It is well known from previous results and simulations
that the behaviour is strikingly different for the case
 $f=1$ than for larger $f$ \cite{FA,Harrowell,ReviewKLG}.
With $f=1$,  a single vacancy can facilitate the flip of any of
its neighbors, therefore
% a single down spin is a {\sl defect} that can diffuse
%on the lattice and therefore
 relaxation does not require cooperative processes
and time scales are proportional to the density of vacancies, as confirmed by numerics.
This least-restricted  FA model is loosely analogous to the 
$s-1$ KA models 
normal hard core lattice gas (which would correspond to KA with $s=0$)
for which a single vacancy can freely
move in an otherwise totally filled lattice: in the FA case,  a vacancy can
``move'' by flipping  a neighboring occupied site, and then flipping to occupied at the original site.
Nevertheless, this analogy cannot be take too far: interesting --- although not cooperative --- behavior does exist for 
$f=1$ FA models. Indeed it has been shown in \cite{Whitelam} 
that in dimension less than $four$ vacancies cannot be considered as
independently diffusing defects (in contrast to the trivial KA model in which single vacancies can move). The interaction between vacancies, in particular their  creation and
annihilation, lead to a reaction-diffusion process that belongs to the
same universality class as directed percolation in three dimensions.
%This leads to a typical relaxation
%time that is proportional to the density of up spins

For more restricted FA models with the facilitating parameter $f>1$
% due to the
%stronger constraints
neither a single facilitating vacancy, nor a finite
cluster of them,  can enable all the other sites to flip if these are initially 
all occupied:  as for KA models, a finite vacancy cluster can never breakthrough
a slab that is infinite in $2d-f+1$ directions and of width two in the other $f-1$ directions. 
Hence, cooperative processes must be 
involved in the low temperature relaxation of these FA models.
The nature of the cooperative mechanism that causes equilibration has been
conjectured by Reiter \cite{Reiter}: it  is very similar to the one 
we discussed for the KA model in \cite{letter} with the correspondence $f\to s+1$. 

Consider for simplicity the two-dimensional case with $f=2$ at temperature $T$ 
and focus on
regions of an infinite lattice that are  frameable ---- in the KA sense with
$s=1$ (see section \ref{d2})---- out to linear  size $\xi (\rho)$, with
$\rho(T)$ defined in (\ref{gibbs}) and $\xi$ as in (\ref{xi2}).
These we refer to as mobile cores.
\cite{letter}
It is easy to check that using moves allowed by FA rules every 
occupation variable inside the core can be
changed to zero.
On a typical line segment of length $\xi (\rho )$ outside the special core regions,  there is likely to be at least one vacancy. The existence of such vacancies  around  most places a core could be, ensure that a single core can diffuse through typical regions of 
the system --- and eventually everywhere.
Reiter conjectured that the diffusion of these {\sl macro--defects}
is the dominant mechanism for relaxation
in the low temperature regime of FA models. From knowledge of the temperature
dependence of $\xi$ and of the spatial density of mobile cores, $\rho_D$,--- both given by those of the corresponding KA model at the same vacancy density, $\rho$, see section \ref{FAdef}} --- together with  the characteristic time $\tau_{D}$, within a core, one can make a prediction for the temperature dependence of the overall 
 relaxation time. Neglecting, as for KA models, the interactions,
annihilation and creation of the macro-defects, the bulk relaxation time is \cite{letter}: 
\begin{equation}\label{}
\tau \approx \frac{\tau_{D}}{\rho_{D}}\quad .
\end{equation}
This is the time after which typical sites will be changed because a
macro-defect has passed by.\\
Reiter assumed that the typical timescale
$\tau_{D}$ for the
motion of a {\sl macro--defect} of size $l$ grows as $\exp Kl$ with
$K$ a positive constant. However, we have found --- analytically and
numerically --- that this is not the case for the KA model: $\tau_D$
increases more slowly  with $l$ as $\exp K\sqrt{l}$
\cite{letter} (see also section \ref{picture}). 
This result should carry over to the FA case since
it is easier for a 
{\sl macro-defect} to move with  FA dynamics  than with KA dynamics, yet the dynamics within a core is dominated by entropic barriers \cite{letter} which will be the same for both models. 
As a consequence of the sub-exponential slowing down within a core, 
the leading contribution to the structural
relaxation timescale $\tau$ is given by the density of defects and not by
their timescale to move. 
The same argument should hold for any $d$, $1<f\leq d$
giving $\tau\propto \rho_D\simeq 
1/\Xi_{d,f-1}(\rho(T))^{d}$, where 
$\Xi_{d,f-1}(\rho)$ is the crossover length of the 
 corresponding KA model and $\rho(T)$ is the equilibrium density of vacancies
 at temperature $T$, see eq. (\ref{gibs}).
%By rewriting the probability of having
%which is proportional to
%$\left(L^{B}\right)^d$, with $L^{B}$ the crossover length for bootstrap percolation.
%Indeed, as we already mentioned in previous section, bootstrap percolation crossover length
% coincides the one for FA model.
For the case $d=2$ $f=2$, we thus have the exact leading low temperature
 behavior of $\log\tau$:

\begin{equation}\label{faeq}
\tau\simeq\exp{-\frac{2 c_\infty}{1-\rho}}
\end{equation}
where $2c_{\infty}\simeq 1.1$ (see equation (\ref{cinf})).
Numerical simulations on a square lattice agree well
with the above prediction \cite{Sellitto}. However one has to be
careful because of strong finite size effects as discussed for 
the KA case. 
And there is another complication that might arise in the FA
case: the fact that the independent defect diffusion assumption fails in the $f=1$ case
for $d<4$ \cite{Whitelam}, suggests it will also for larger $f$. However if, as for $f=1$, the fluctuations in $d<4$ just change
(slightly) the dynamical exponent that relates time and length, then the scaling of
$\tau $ with the density for  the $d=2$ $f=2$ case would be the same as in (\ref{faeq})  
but perhaps with a (slightly) renormalized constant $c_{\infty }$.  [In the more
cooperative cases for which there are at least two iterated exponentials,  a
change in the dynamical exponent may not even change the leading
asymptotic behavior of the relaxation time,  but this needs a more careful investigation.]

\section{\bf Conclusions}
\label{conclusions}

We have analyzed in this paper a class of kinetically constrained 
lattice models which had previously been studied primarily via numerical simulations.
For the KA models on hypercubic lattices, we have found that the behavior in
infinite systems, in contrast to what was suggested by the simulations, is
simple: either the system is non-ergodic for all densities because of the
existence of finite frozen clusters, or it is ergodic at all
densities.  Nevertheless, in the latter cases  dynamics at high densities is intrinsically collective and thus non-trivial.   We found that there are two length scales that characterize this collective dynamics: The smaller one, $\xi$, is the linear dimension of a minimal set of vacancies that is {\it typically} mobile in a system of density $\rho$; 
such regions of size $\xi^d$ are the mobile cores whose motion allows
rearrangements of other regions.   The longer length scale, $\Xi$, is the
typical spacing between mobile cores. Therefore $\Xi$ is 
the crossover length that determines whether finite-size systems will
typically be almost ergodic, with one irreducible set dominating configuration
space,  or have their configuration space broken up into many large
sets. Furthermore, the same length controls the high density behaviour of 
relaxation times, which scale as the density of mobile cores, therefore as $1/\Xi^d$. The manner in which these lengths diverge as the density approaches unity depends on the dimension and the constraint parameter, $s$.  But, although we have not discussed it thus far, there is considerable universality.

\subsection{Universality}

For simple lattices other than hypercubic, the behavior can be guessed by analogy with the hypercubic cases.  

If there are no finite frozen clusters, but infinite frozen configurations do exist, then the minimum dimension of frozen structures will determine the behavior. If the minimal frozen structures  are $d-1$ dimensional, then the frames of vacancies used to show irreducibility will be one-dimensional and the core size will grow with density as, 
\be
\xi(\rho)\sim\frac{\ln[1/(1-\rho)]}{(1-\rho)^{p/(d-1)}}
\ee
where $p$ is the number of nearest neighbor vacancies needed
on each side to enable  a frame to grow. The crossover scale grows
roughly as $\Xi\sim e^\xi$.
This least-restricted collective behavior obtains for $s=1$ on hypercubic lattices in all dimensions, and also for, e.g., two-dimensional triangular lattices with
$s=2$ ($s=1$ has mobile pairs of vacancies); and three-dimensional fcc
lattices (close-packed) with $s=3$ ($s=1$ or $2$ models have small
mobile clusters of vacancies). In all these cases $p=1$.

If the   smallest frozen structures are $d-2$ dimensional, framing by 
two-dimensional planes is needed: this gives rise to 
\be
\xi \sim \exp\left(\frac{c}{(1-\rho)^{p/(d-2)}}\right)
\ee
This form cannot occur in two dimensions. In three dimensions, it obtains for
cubic-lattices with $s=2$, as well as for fcc lattices, for $s=4$ and $s=5$, these two cases only differing by the numerical coefficient.  

In three dimensions, the above  are the only possibilities in the same general class that we have been studying, but in higher dimensions, the iterated exponential forms of (\ref{CLds}) can occur. 
As we have shown, on Bethe lattices --- expected to be loosely related to some high dimensional limit --- an actual dynamical transition  at a non-trivial density, $\rho_c$,
does  occur for KA models.

This immediately suggests a  crucial question: Do some
finite dimensional models exist with this type of dynamical transition? Two
candidates in the literature are infinitely thin hard rods on a
lattice \cite{HardRods} and a space filling self-avoiding chain on a
lattice \cite{Grosberg}. We leave further exploration for the  future. 

But the issue of {\it apparent transitions} does already arise in the simple KA models.  As mentioned at the end of section \ref{picture} 
dramatic slowing down in finite dimensions
can occur near an almost sharp {\sl avoided transition} associated with a  
crossover from  non-cooperative to cooperative dynamics.  In particular, some high dimensional lattices --- including already
some three dimensional lattices --- are approximately tree-like out to some
number of neighbors; if the dimension increases while the
coordination number remains fixed, such tree-like local structure suggests that a ghost of the Bethe
lattice transition
 with the same local structure will become more and more apparent. More generally, it is likely that in the limit of high dimensions, models that have low-dimensional frozen structures --- such as $s=d-1$ on hypercubic lattices which has frozen one-dimensional bars --- will, as the density increases,  have increasingly rapid  crossovers that  appear more and more like an actual transition.

 In this paper we have studied Bethe lattices with loops 
that interpolate between the finite dimensional lattice
and the standard Bethe lattice. In this case there is always a transition 
although it is pushed to a density that approaches one when the finite-dimensional regions become large. From the other direction, it would be  
interesting to analyze a Kac-like interpolation for a fixed
finite dimensional lattice with longer and longer range connections which converges in a certain limit to a Bethe
lattice. In this case the transition will presumably never occur, but the cross-over
will become sharper and sharper near the critical density on the Bethe lattice.

\subsection{Comparison with mean-field approaches}

We have found that the generic  behavior of KA models on Bethe lattices  (exceptions
being $s=k-1$), has a transition with features similar to both first order and critical
transitions. Although the density of frozen particles jumps
discontinuously at the critical density, there are precursor effects
as the transition is approached from the frozen phase and, at least
for some quantities, also from the ergodic phase. In particular, we have seen that  divergence of characteristic
length and time scales is apparent in the dynamic density-density
correlations and the related dynamical susceptibility.

This mixed behavior, as well as several more specific features, is similar to that found in certain mean-field-like theoretical analyses conjectured to be relevant for glass transitions.
In particular,  spin models with  {\it quenched random} interactions 
that couple
collections of $p>2$ spins 
have been studied \cite{CuKu,ReviewOE}. It turns out that
the infinite-range versions of such random p-spin models 
 have a real
thermodynamic transition at a critical temperature, however interesting
dynamical behavior commences sharply at a higher temperature, $T_D$,
without singularities in static properties. This is interpreted as the
onset of ergodicity breaking: below $T_D$ the equilibrium measure is
fractured in exponentially many
metastable ``states" (``TAP" states which are locally stable solutions
of the self-consistent mean-field equations); the large number of these give rise to a
configurational entropy density, $S_c$,  that  jumps to a non-zero
value at $T_D$.  As the temperature is lowered further, $S_c$
decreases, vanishing at the thermodynamic transition.  
Due to the presence of quenched random disorder p-spin models
 would seem to have little to do with glasses which must generate their own
 randomness, but  it has been shown that some mean-field approximations to
 non-random models, in particular the  mode-coupling approximation,
 \cite{Gotze} give rise to self consistent integral equations with very
 similar structure to the one of this models\cite{ReviewOE}. In
 particular, mode coupling equations give rise to an analogous 
ergodicity breaking transition at a finite.
We find that the same properties are shared 
by KA models on Bethe lattices,
for which at a finite critical density the configuration space 
is broken into many
ergodic components and the configurational entropy jumps to a non-zero value
and then decreases at higher densities (here the thermodynamics is trivial and, hence, $S_{C}$ equals zero only at unit density).
In addition, the autocorrelation function, $C(t)$, and the associated
susceptibility, $\chi_4(t)$, which we computed numerically  for Bethe lattices
with $k=3$ $s=1$, behave similarly to those found for the infinite range
random $p$-spin models and in the Mode Coupling Theory \cite{KT,FranzDH, BiroliBouchaud}.
Indeed, even the off-equilibrium aging behavior of KCM on Bethe lattices is quite similar
to the one obtained for mean-field quenched random systems \cite{SellittoBiroliToninelli}.

While tempting to conclude that these behaviors are all manifestations of the
same physical phenomena, one must be careful. In particular the
dynamical transition in KA models on Bethe lattices is a
reducible/irreducible transition whereas for p-spin models the irreducibility
is guaranteed and the breaking of the ergodicity is due to the
thermodynamic limit. Nevertheless, to explore whether, and in what sense, some of the mean-field approximations might capture some of the essential physics of more realistic finite dimensional models, the possible connections to the KA models are worth exploring further.

\subsection{Experimental issues and prospects}

Kinetically constrained lattice gas models, while they may capture some of the important features of glass transitions --- or almost transitions ---, clearly have major flaws.
First, in any real system, rates of local processes cannot strictly vanish. Thus there will also be some constraint-violating processes. 
Second, while lattice models are often good approximations for static
properties, for dynamics of systems in which the geometry and/or
bonding is believed to play an important role, it is dangerous to
appeal to universality arguments for the validity of lattice
models. This is especially true if the length scales of the important
processes never get very large, as is believed to be the case for
glasses \cite{Erdiger}. 

Both of these issues merit much further attention.  In particular, one should at a minimum consider the effects of a low rate of constraint-violating particle motions within the lattice model framework.  In practice, the neglect of certain processes may be a good approximation on a wide range of time scales.  For example, if there are chemical bonds with energies much larger than $k_B T$, then the ratio of the rates of processes between, e.g., those that require breaking of two bonds and those that require breaking three or more, may be very small.  A --- and arguably {\it the} --- fundamental problem of dynamics near glass transitions is the large and rapidly-onsetting increases in the effective activation barriers. Thus looking for an explanation that involves a rapid crossover from domination by a relatively fast process to freezing out of this and the ensuing domination of relaxation by a much slower process, is surely a productive direction. 

The issue of continuum versus lattice models is a trickier one.  If
geometrical constraints really dominate near glass transitions, then
one needs to consider how to model the interplay between energetic and
entropic barriers, even if the former only arise from repulsive
interactions that are not entirely hard-core.  In principle, from
simulations or from computations of more realistic continuum models of
molecular interactions, one may be able to extract effective rates for
various local processes (see \cite{ChandlerGarrahanKLG} for an investigation 
in this direction). If temperature changes dominate over
density changes \cite{Alba} the effective density parameter in a
lattice approximation will not be simply related 
to the actual density because of soft-core effects. Nevertheless, if free-volume ideas of glass transitions are relevant, one should be able to work in terms of some effective density --- with  ``vacancies", or at least mobile ones, representing the free volume. Because of local expansion and contraction, however, the free volume is in general not conserved. One might expect the particle conservation to play a crucial role in KA  models.  In fact, this is not the case, as the  non-conserving Fredrickson-Anderson model 
shows very similar behavior to KA models.  Presumably, a better approximation would conserve (effective) density most of the time, but not always.

We leave these issues and any attempts at direct comparisons with experiments, for future papers.

\section{Acknowledgments}

This work has been supported in part by the National Science
Foundation via grants DMR-9976621 and DMR-0229243. The authors would
like to thank L. Berthier, J.-P. Bouchaud, and D. Chandler
for useful discussions. 

\addcontentsline{toc}{chapter}{{\bfseries Bibliography}}

\end{document}